\documentclass[preprint,12pt]{elsarticle}
\usepackage[top=0.8in, bottom=0.8in, left=0.9in, right=0.9in]{geometry}
\usepackage{amssymb}
\usepackage{amsthm}
\usepackage{amsmath}
\usepackage{enumitem}
\usepackage{bm}
\usepackage{graphicx}
\usepackage{graphics}
\usepackage{subfigure}
\usepackage{subcaption}
\usepackage{tabularx,caption}
\usepackage{float}
\usepackage{color}
\usepackage{multirow}
\usepackage{booktabs}
\usepackage{diagbox}
\usepackage{lineno}
\biboptions{sort&compress}
\usepackage[unicode]{hyperref}

\begin{document}

\begin{frontmatter}

\title{Monte Carlo physics-informed neural networks for inverse multiscale heat conduction problems via the phonon Boltzmann transport equation}

\author[hust]{Qingyi Lin}
\author[hdu]{Chuang Zhang}
\author[hust]{Xuhui Meng \fnref{1}}
\author[hust]{Zhaoli Guo}

\address[hust]{Institute of Interdisciplinary Research for Mathematics and Applied Science, School of Mathematics and Statistics, Huazhong University of Science and Technology, Wuhan 430074, China}
\address[hdu]{Institute of Fundamental and Transdisciplinary Research, Zhejiang University, Hangzhou 310058, China}

\fntext[1]{Corresponding author: xuhui\_meng@hust.edu.cn (Xuhui Meng).}

\begin{abstract}
Inferring thermal fields and thermophysical properties from limited measurements is a fundamental challenge in micro- and nanoscale heat conduction, where the classical Fourier law breaks down and the phonon Boltzmann transport equation (BTE) must be employed to capture non-diffusive transport effects.
In this work, we extend the Monte Carlo physics-informed neural networks (MC-PINNs), originally developed for forward phonon BTE problems [{\emph{J. Comput. Phys. 542, 114364, 2025}}], to solve inverse multiscale heat conduction problems.
Two representative classes of inverse problems are considered: (i) reconstructing the full thermal field from sparse interior temperature measurements when boundary conditions are unknown, and (ii) simultaneously inferring the unknown relaxation time, which characterizes the transport regime, together with the thermal field.
Problem-specific MC-PINNs architectures and training strategies are designed for each class.
The mesh-free Monte Carlo sampling strategy, inherited from the forward formulation, enables a unified treatment across diffusive, transitional, and ballistic transport regimes without requiring a priori knowledge of the relaxation time.
The proposed method is evaluated on quasi-one-dimensional, quasi-two-dimensional, and three-dimensional benchmark problems covering a wide range of Knudsen numbers, as well as on a realistic 3D fin field-effect transistor (FinFET) structure.
Results demonstrate that MC-PINNs consistently outperform purely data-driven deep neural networks, particularly in the sparse-data regime, and can accurately infer spatially uniform relaxation times.
For spatially varying relaxation times, the inferred distributions capture the dominant thermal response, and numerical simulations using the recovered parameters reproduce the macroscopic fields with good accuracy.
These findings establish MC-PINNs as an effective and physically consistent framework for inverse thermal analysis at micro- and nanoscales.
\end{abstract}


\end{frontmatter}

\section{Introduction}

Understanding and controlling heat conduction at micro- and nanoscales is essential for a wide range of applications, including thermal management of microelectronic devices, defect detection via thermography, and design of nanoscale materials~\cite{liu2007thermography,aldave2013review,christofferson2008microscale,li2022highly}.
While the classical Fourier law provides an accurate description of heat conduction in most macroscopic systems, its validity breaks down when the characteristic length or time scale approaches the mean free path or relaxation time of energy carriers~\cite{benenti2023non,chen2021non,wang2011non}.
Under such conditions, encountered at cryogenic temperatures, in nanoscale structures, or under ultrafast laser heating, non-diffusive and ballistic transport effects become significant, and the phonon Boltzmann transport equation (BTE) provides a more fundamental description of the heat conduction process~\cite{wang2011non,kaiser2017thermal,mcgaughey2019phonon}.

Conventional numerical methods for solving the phonon BTE, such as the discrete ordinates method and Monte Carlo simulations, generally require complete knowledge of the thermophysical properties as well as precise boundary and initial conditions~\cite{guo2016lattice,zhang2017unified,zhang2023acceleration}.
In many practical scenarios, however, such information is incomplete or unavailable due to complex geometries, interfacial effects, or experimental limitations~\cite{ames1965nonlinear,tu2023computational}.
At the same time, partial temperature measurements inside the domain can often be acquired through advanced sensing and imaging techniques~\cite{reihani2021quantitative}.
This combination of incomplete governing information and sparse observational data naturally leads to inverse heat conduction problems, in which one seeks to infer the full thermal field and/or unknown physical parameters from limited measurements.

Classical approaches to field reconstruction from sparse data, such as interpolation, example-based internal learning, and neighbor embedding~\cite{fadnavis2014image,kim2008example}, typically rely on assumptions of spatial smoothness or local correlations, which limits their ability to resolve sharp thermal gradients or capture nonlocal transport effects characteristic of the BTE.
In recent years, deep learning methods have emerged as powerful tools for tackling such inverse problems by learning complex nonlinear mappings from data~\cite{aggarwal2018modl,bai2020deep,arridge2019solving}.
Convolutional neural networks and generative adversarial networks, for instance, have been successfully applied to super-resolution and high-fidelity reconstruction across various physical domains~\cite{mccann2017convolutional,ongie2020deep,jin2017deep,wei2019physics,xie2018tempogan,liu2020deep,fukami2019super}.
However, purely data-driven models typically require large, high-resolution training datasets that are rarely available in practical thermal measurements, and the absence of explicit physical constraints may lead to predictions that violate the underlying transport physics.

Physics-informed neural networks (PINNs)~\cite{raissi2019physics,cai2021physics} offer an attractive alternative by embedding governing equations directly into the training objective through automatic differentiation~\cite{baydin2018automatic}.
This enables the inference of physically consistent solutions from sparse data without requiring large training sets, making PINNs particularly well-suited for reconstructing physical fields and identifying unknown parameters from limited or noisy measurements~\cite{chen2020physics,jagtap2022physics,bian2023high,gao2021super,cai2021flow}.
Recent studies have demonstrated the effectiveness of PINNs for thermal field reconstruction and parameter identification~\cite{cai2021physics2,zhang2022multi,liu2022temperature,li2024solving,tang2025physics,koric2023data}.
Nevertheless, most existing PINN-based thermal studies have focused on macroscopic heat conduction governed by Fourier's law, and multiscale problems that require the phonon BTE remain largely unexplored in the inverse setting.

The application of PINNs to the phonon BTE has so far been limited to forward modeling.
Li {\sl et al.}~\cite{li2021physics,li2022physics} demonstrated that PINNs can achieve satisfactory accuracy for forward phonon BTE problems across transport regimes ranging from diffusive to ballistic.
However, their formulation relies on a priori discretization of the solid angular space, which requires a large number of angular collocation points to resolve the ballistic and near-ballistic regimes, limiting scalability to high-dimensional problems.
More critically, when thermophysical properties are unknown, as is typically the case in inverse problems, an appropriate angular discretization cannot be prescribed in advance, because the transport regime is not known a priori.

To overcome these limitations, Lin {\sl et al.} recently proposed Monte Carlo PINNs (MC-PINNs)~\cite{lin2025monte} for the forward phonon BTE.
The method employs a two-step Monte Carlo sampling strategy, randomly drawing collocation points in both physical and angular spaces at each training iteration.
This mesh-free approach removes the need for predefined spatial or angular grids, enables efficient exploration of the computational domain across all transport regimes, and reduces computational cost compared with grid-based PINN formulations.
These properties make MC-PINNs a natural foundation for inverse phonon BTE problems, where the transport regime is unknown and a flexible, grid-free treatment is essential.

In the present work, we extend the MC-PINNs to inverse multiscale heat conduction problems governed by the phonon BTE.
Two representative classes of inverse problems are investigated, each reflecting a distinct source of ill-posedness commonly encountered in micro- and nanoscale thermal measurements:
\begin{enumerate}[label=(\arabic*)]
    \item \textbf{Thermal field reconstruction with known relaxation time.}
    The governing equation is fully known but boundary and/or initial conditions are incomplete. Given sparse interior temperature measurements, the objective is to reconstruct the full thermal field over the entire domain.
    \item \textbf{Simultaneous inference of the thermal field and unknown relaxation time.}
    The boundary and initial conditions are prescribed, but the relaxation time $\tau$ is unknown. Two sub-cases are examined: (2a) $\tau$ is spatially uniform but unknown, and (2b) $\tau$ is an unknown spatially varying function $\tau(\boldsymbol{x})$.
\end{enumerate}
For each problem class, we develop tailored neural network architectures and training strategies within the unified MC-PINNs framework.
The framework's effectiveness is systematically evaluated on quasi-one-dimensional (quasi-1D), quasi-two-dimensional (quasi-2D), and three-dimensional (3D) test cases spanning diffusive to ballistic transport regimes, including a realistic 3D FinFET structure representative of state-of-the-art microelectronic devices.

The remainder of the paper is organized as follows.
Section~\ref{sec:methodology} introduces the phonon BTE, formulates the two classes of inverse problems, and presents the MC-PINNs for solving them.
Section~\ref{sec:results} presents comprehensive numerical experiments for thermal field reconstruction and relaxation time inference.
Section~\ref{sec:conclusion} summarizes the findings and discusses directions for future work.

\section{Methodology}\label{sec:methodology}

\subsection{The phonon Boltzmann transport equation and inverse problem formulation}\label{sec:bte}

We adopt the gray-model approximation~\cite{kaiser2017thermal} to describe multiscale heat conduction in solids, following the formulation in Ref.~\cite{lin2025monte}.
Under this approximation, phonon dispersion and polarization are neglected, and all phonons are assumed to share the same group speed.
The dimensionless gray phonon BTE reads
\begin{equation}\label{eq:bte}
\partial_t e^{\prime\prime}+\boldsymbol{v}\cdot\nabla e^{\prime\prime}=\frac{1}{\tau}[e^{eq}-e^{\prime\prime}] + \dot{S},
\end{equation}
where $e^{\prime\prime}$ denotes the phonon energy density per unit solid angle, and $\boldsymbol{v} = v_g\boldsymbol{s}$ is the group velocity with group speed $v_g$ and unit directional vector $\boldsymbol{s}=(\cos\theta, \sin\theta\cos\phi, \sin\theta\sin\phi)$.
Here, $\theta\in[0,\pi]$ and $\phi\in[0,2\pi)$ are the polar and azimuthal angles, respectively; $\tau$ is the relaxation time; and $\dot{S}$ is the volumetric heat source~\cite{guo2016lattice,zhang2017unified,zhang2023acceleration}.
The equilibrium phonon energy density $e^{eq}$ is given by
\begin{equation}\label{eq:eeq}
e^{eq}(t, \boldsymbol{x},\boldsymbol{s})=\frac{C_VT}{4\pi},
\end{equation}
where $C_V$ is the volumetric heat capacity.
In this work, the volumetric heat capacity and group velocity are normalized by their respective reference values, yielding dimensionless values $C_V=1$ and $v_g=1$.
The macroscopic temperature $T$ and heat flux vector $\boldsymbol{q}$ are obtained by taking angular moments of $e^{\prime\prime}$:
\begin{align}\label{eq:macro}
T = \frac{\int e^{\prime\prime} d\Omega}{C_V}, \quad \bm{q} = \int  \bm{s} e^{\prime\prime} d\Omega.
\end{align}

The thermalization boundary condition is applied to close Eq.~\eqref{eq:bte}~\cite{fryer2014moment}.
Under this condition, phonons incident on a boundary are absorbed, while emitted phonons are assumed to be in equilibrium with the wall temperature, leading to the Dirichlet condition
\begin{equation}\label{eq:bc}
e^{\prime\prime}(t,\boldsymbol{x}_w, \boldsymbol{s}) =\frac{C_V T_w}{4\pi},\quad \boldsymbol{s}\cdot \boldsymbol{n}>0,
\end{equation}
where $\boldsymbol{x}_w$ denotes a boundary point, $\boldsymbol{n}$ is the outward unit normal, and $T_w$ is the prescribed wall temperature.
The initial condition for a system initially at thermal equilibrium with temperature $T_0$ is~\cite{guo2016discrete}
\begin{equation}\label{eq:ic}
e^{\prime\prime}(t_0,\boldsymbol{x}, \boldsymbol{s}) =\frac{C_V T_0}{4\pi}.
\end{equation}

The transport regime is characterized by the Knudsen number $\mathrm{Kn} = \lambda/L_0$, where $\lambda=v_g\tau$ is the phonon mean free path and $L_0$ is the characteristic length scale.
In the diffusive limit ($\mathrm{Kn}\ll 1$), the BTE reduces to Fourier's law.
As $\mathrm{Kn}$ increases, nonlocal and ballistic effects become increasingly important, and the system enters the transitional and ultimately the ballistic regime ($\mathrm{Kn}>10$).
Since $L_0$ and $v_g$ are prescribed in each test case, the relaxation time $\tau$ serves as a proxy for $\mathrm{Kn}$, with different values of $\tau$ corresponding to distinct transport regimes.

We now formulate the two classes of inverse problems addressed in this work.
In contrast to the forward setting considered in Ref.~\cite{lin2025monte}, where all thermophysical properties, boundary conditions, and initial conditions are fully specified, inverse problems are characterized by incomplete information about the governing equation or its auxiliary conditions.
Such situations frequently arise in practice due to inaccessible boundaries, uncertainties in material properties, or limitations of experimental characterization techniques~\cite{cahill2014nanoscale}.
\begin{enumerate}[label=(\arabic*)]
    \item \textbf{Thermal field reconstruction with known relaxation time.}
    The relaxation time $\tau$ is known, but the boundary and/or initial conditions are missing. Partial temperature measurements from sensors distributed within the domain are available, and the objective is to reconstruct the full thermal field.
    This setting represents situations where the governing physics is well characterized but boundary conditions are difficult to specify, for instance due to complex device geometries or inaccessible interfaces.

    \item \textbf{Simultaneous inference of the thermal field and unknown relaxation time.}
    The boundary and initial conditions are prescribed, but the relaxation time $\tau$, which determines the transport regime, is unknown.
    Given sparse temperature and/or heat flux measurements, the goal is to jointly reconstruct the thermal field and identify $\tau$.
    Two sub-cases are distinguished: (a) $\tau$ is spatially uniform but unknown; and (b) $\tau$ is an unknown spatially varying function $\tau(\boldsymbol{x})$.
    This setting is representative of material characterization problems where the thermophysical properties of the sample are not known a priori.
\end{enumerate}
These two problem classes represent different sources of ill-posedness.
In Problem (1), the governing equation is known but the missing boundary conditions prevent a direct forward solution.
In Problem (2), the boundary conditions are available but an unknown coefficient in the governing equation must be inferred from sparse observations together with the solution field.
This structural difference motivates the distinct solution strategies developed below.

\subsection{MC-PINNs for inverse multiscale heat conduction}\label{sec:mcpinns}

We extend the MC-PINNs framework developed in Ref.~\cite{lin2025monte} to the inverse problems formulated above.
We first briefly summarize the forward MC-PINNs formulation, which serves as the foundation, and then describe how it is adapted to each class of inverse problems.

\subsubsection{Forward MC-PINNs formulation}

In the forward setting, the solution to the phonon BTE is approximated by a deep neural network $e_{\mathrm{NN}}(t,\boldsymbol{x},\boldsymbol{s}; \bm{\theta})$ with trainable parameters $\bm{\theta}$, which takes the temporal, spatial, and angular coordinates as inputs and predicts the phonon energy density.
The network is trained by minimizing a composite loss function that enforces the governing equation together with the prescribed boundary and initial conditions:
\begin{equation}\label{eq:loss_fwd}
    \mathcal{L}_{\mathrm{fwd}} = \omega_{Eq}\mathcal{L}_{Eq} + \omega_{\mathrm{BC}}\mathcal{L}_{\mathrm{BC}} + \omega_{\mathrm{IC}}\mathcal{L}_{\mathrm{IC}},
\end{equation}
where the individual terms are defined as
\begin{equation}\label{eq:loss_fwd_terms}
    \begin{aligned}
       \mathcal{L}_{Eq}&=\frac{1}{N}\sum^{N}_{i=1}\left|R_i(t, \boldsymbol{x}, \boldsymbol{s})\right|^2, \\
       \mathcal{L}_{\mathrm{BC}}&=\frac{1}{N_{\mathrm{BC}}}\sum^{N_{\mathrm{BC}}}_{j=1}\left|e_{\mathrm{NN},j}(t, \boldsymbol{x}_w, \boldsymbol{s})-e^{eq}_j(t, \boldsymbol{x}_w, \boldsymbol{s})\right|^2,\\
       \mathcal{L}_{\mathrm{IC}}&=\frac{1}{N_{\mathrm{IC}}}\sum^{N_{\mathrm{IC}}}_{k=1}\left|e_{\mathrm{NN},k}(t_0, \boldsymbol{x}, \boldsymbol{s}) - e_k^{eq}(t_0, \boldsymbol{x}, \boldsymbol{s})\right|^2,
    \end{aligned}
\end{equation}
with the PDE residual
\begin{equation}\label{eq:residual_fwd}
R=\partial_t e_{\mathrm{NN}}+\boldsymbol{v}\cdot\nabla e_{\mathrm{NN}} - \frac{1}{\tau}(e^{eq}- e_{\mathrm{NN}}) - \dot{S}.
\end{equation}
Here, $N$, $N_{\mathrm{BC}}$, and $N_{\mathrm{IC}}$ denote the numbers of residual, boundary, and initial collocation points, and $\omega_{Eq}$, $\omega_{\mathrm{BC}}$, and $\omega_{\mathrm{IC}}$ are user-specified loss weights.
For steady-state problems, the initial condition term $\mathcal{L}_{\mathrm{IC}}$ is omitted.

A defining feature of MC-PINNs is the two-step Monte Carlo sampling strategy used to construct training points at each iteration.
In the first step, $B_{t,\boldsymbol{x}}$ points are randomly drawn in the temporal-spatial domain; in the second step, $B_{\boldsymbol{s}}$ points are independently drawn in the solid angular domain.
The training points are then formed as the Cartesian product of these two sets.
Since the collocation points are regenerated at each training step, the angular space is explored stochastically over the course of optimization, ensuring sufficient directional coverage without requiring a fixed angular grid.
This sampling strategy eliminates the need for predefined discretizations, naturally handles complex geometries, mitigates the curse of dimensionality, and also achieves a unified, mesh-free treatment across all transport regimes \cite{lin2025monte}.

\subsubsection{Extension to inverse problems}

To address the inverse problems formulated in Sec.~\ref{sec:bte}, the forward MC-PINNs framework is augmented with a data-fidelity term $\mathcal{L}_{\Phi}$ that incorporates available measurements.
The composite loss function becomes
\begin{equation}\label{eq:loss}
    \mathcal{L} = \omega_\Phi\mathcal{L}_{\Phi} + \omega_{Eq}\mathcal{L}_{Eq} + \omega_{\mathrm{BC}}\mathcal{L}_{\mathrm{BC}} + \omega_{\mathrm{IC}}\mathcal{L}_{\mathrm{IC}},
\end{equation}
with
\begin{equation}\label{eq:data_loss}
    \mathcal{L}_{\Phi}=\frac{1}{N_{\mathrm{sen}}}\sum^{N_{\mathrm{sen}}}_{m=1}\left|\Phi_{\mathrm{sen},m}(t, \boldsymbol{x}) - \Phi_{\mathrm{pred},m}(t, \boldsymbol{x})\right|^2,
\end{equation}
where $\Phi$ denotes the observable macroscopic quantities (temperature $T$ and/or heat flux $\boldsymbol{q}$), $N_{\mathrm{sen}}$ is the number of sensor locations, and $\Phi_{\mathrm{sen},m}$ and $\Phi_{\mathrm{pred},m}$ are the measured and predicted values, respectively.
The predicted macroscopic quantities are computed from the network output $e_{\mathrm{NN}}$ via the moment integrals in Eq.~\eqref{eq:macro}, with the angular integrals evaluated by Monte Carlo quadrature using the sampled angular points.
The loss terms $\mathcal{L}_{Eq}$, $\mathcal{L}_{\mathrm{BC}}$, and $\mathcal{L}_{\mathrm{IC}}$ retain the definitions in Eq.~\eqref{eq:loss_fwd_terms}; any term for which the corresponding information is unavailable is simply dropped.
Unless otherwise noted, unit loss weights ($\omega_{\Phi}=\omega_{Eq}=\omega_{\mathrm{BC}}=\omega_{\mathrm{IC}}=1$) are used throughout, as they are found to yield satisfactory accuracy empirically.

\begin{figure}
    \centering
    \includegraphics[width=1.0\textwidth]{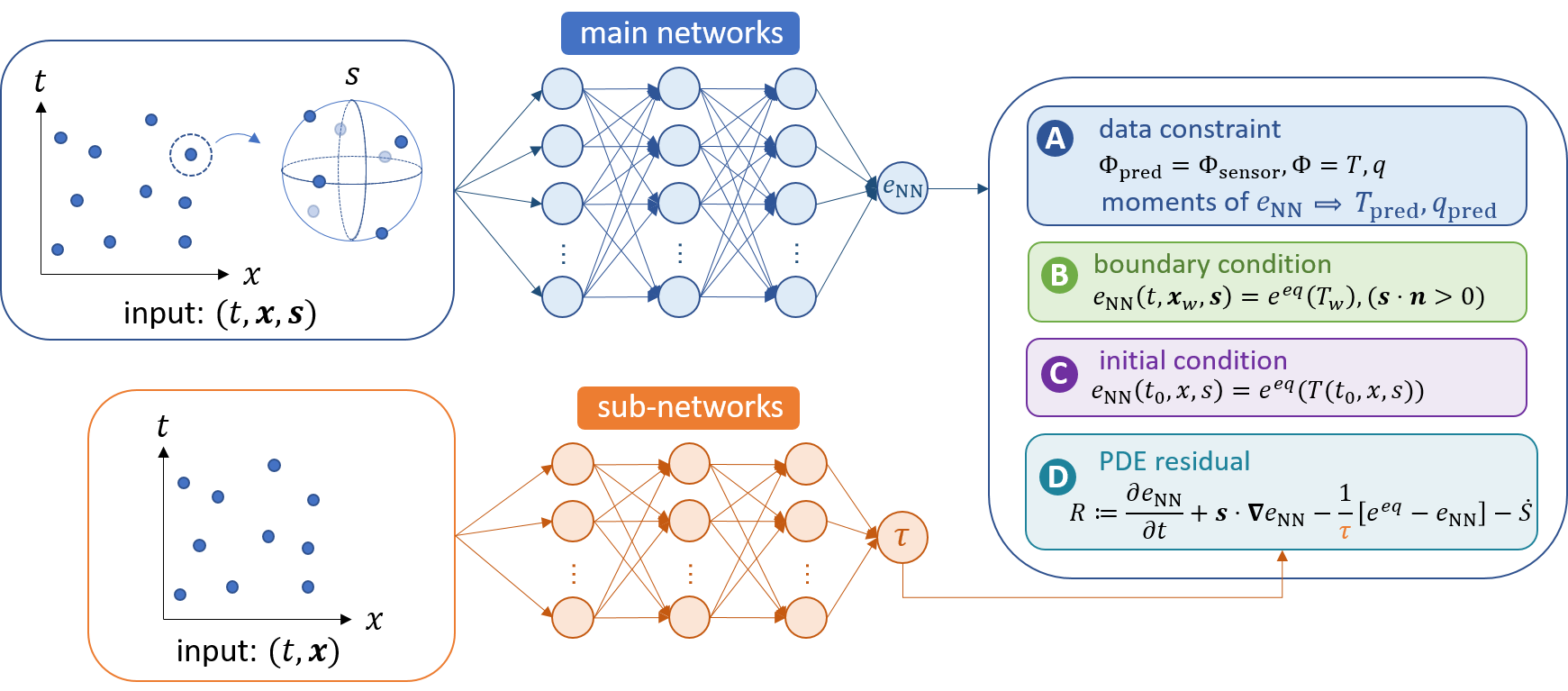}
    \caption{Schematic of the MC-PINNs framework for solving inverse problems governed by the phonon BTE.
    The main network takes $(t, \boldsymbol{x}, \boldsymbol{s})$ as inputs and outputs the phonon energy density $e_{\mathrm{NN}}$. An optional sub-network is employed to approximate spatially varying relaxation times when required.
    The loss function combines observational data with the available physical constraints, including the governing equation and boundary/initial conditions.}
    \label{fig:PINNs}
\end{figure}

The solution strategy is tailored to each inverse problem class, as illustrated in Fig.~\ref{fig:PINNs}.

\textbf{Problem (1) -- Unknown boundary conditions.}
The main network $e_{\mathrm{NN}}(t,\boldsymbol{x},\boldsymbol{s})$ approximates the energy density distribution as in the forward setting.
Since boundary conditions are unavailable, $\mathcal{L}_{\mathrm{BC}}$ is omitted, and the loss function consists of the PDE residual, the initial condition (for transient problems), and the data-fidelity term evaluated at the sensor locations.
The known relaxation time $\tau$ is directly used in the scattering term of the residual.

\textbf{Problem (2a) -- Spatially uniform unknown $\tau$.}
The main network architecture remains the same, but the unknown relaxation time $\tau$ is treated as a trainable parameter and optimized jointly with the network weights.
The boundary/initial conditions are fully prescribed, so the loss function includes all four terms in Eq.~\eqref{eq:loss}, with $\mathcal{L}_{\Phi}$ providing the additional constraint needed to identify $\tau$.

\textbf{Problem (2b) -- Spatially varying unknown $\tau$.}
An auxiliary sub-network is introduced to approximate the spatially varying relaxation time $\tau(t,\boldsymbol{x})$.
This sub-network takes $(t,\boldsymbol{x})$ as inputs and outputs $\tau$, with an exponential activation function applied at the output layer to enforce positivity.
The inferred $\tau$ is dynamically embedded into the scattering term of the main network's PDE residual, and both networks are trained jointly.

The mesh-free Monte Carlo sampling strategy is especially useful in the inverse setting.
Since $\tau$ is either unknown or its value cannot be used to determine an appropriate angular discretization a priori, the grid-free nature of MC-PINNs ensures that the angular space is adequately explored regardless of the transport regime.
This unified treatment across diffusive, transitional, and ballistic conditions is an advantage over grid-based PINNs or conventional deterministic numerical methods, which generally require the angular discretization to be chosen according to the expected Knudsen number.

\section{Results and Discussion}\label{sec:results}

This section presents a comprehensive evaluation of the proposed MC-PINNs framework on the two classes of inverse problems introduced in Sec.~\ref{sec:methodology}.
In all test cases, reference solutions are obtained by solving the corresponding forward phonon BTE problems under complete boundary/initial conditions using the numerical methods of Refs.~\cite{zhang2023acceleration,zhang2019implicit,zhang2025effects}.
Sensor measurements used for training are mimicked by sampling from these reference solutions at specified locations.
The network architectures, activation functions, and detailed computational settings for each case are provided in~\ref{sec:appendix_c}. The prediction accuracy is quantified by the relative $L_2$ error,
\begin{equation}\label{eq:l2}
E_{\Phi} = \frac{\Vert \Phi_{\mathrm{ref}} - \Phi_{\mathrm{pred}} \Vert_2}{\Vert \Phi_{\mathrm{ref}}\Vert_2},\quad \Phi={T, q, \mathrm{or}\,\tau},
\end{equation}
where $\Phi_{\mathrm{pred}}$ denotes the prediction (from either MC-PINNs or a purely data-driven deep neural network, DNN, used for comparison) and $\Phi_{\mathrm{ref}}$ is the reference solution.

\subsection{Thermal field reconstruction with known relaxation time}\label{sec:results_known_tau}

We first consider Problem (1), in which the relaxation time $\tau$ is known but the boundary conditions are unavailable, and the task is to reconstruct the full thermal field from sparse interior temperature measurements.
Two representative configurations are studied: quasi-2D steady-state heat conduction in a square domain, and 3D heat conduction in a bulk FinFET structure.
For each case, MC-PINNs are compared with purely data-driven DNNs trained on the same sensor measurements, to assess the effect of embedding physical constraints.

\subsubsection{Quasi-2D heat conduction in a square domain}

We begin with quasi-2D steady-state heat conduction in a square domain of side length $L=1$.
Four relaxation times, i.e., $\tau=0.01$, $0.1$, $1.0$, and $10.0$, are examined, spanning transport regimes from near-diffusive to ballistic.
For each case, $B_{\boldsymbol{x}}=100$ spatial points and $B_{\boldsymbol{s}}=100$ angular points are randomly sampled to construct the training points used in the physics-informed loss at each training step.
In this particular case, both MC-PINNs and DNNs are trained using $N_{\mathrm{sen}}=20$ sensor measurements randomly sampled from the reference solutions~\cite{zhang2023acceleration}.

\begin{figure}[H]
\centering
\subfigure[]{\includegraphics[width=0.24\textwidth]{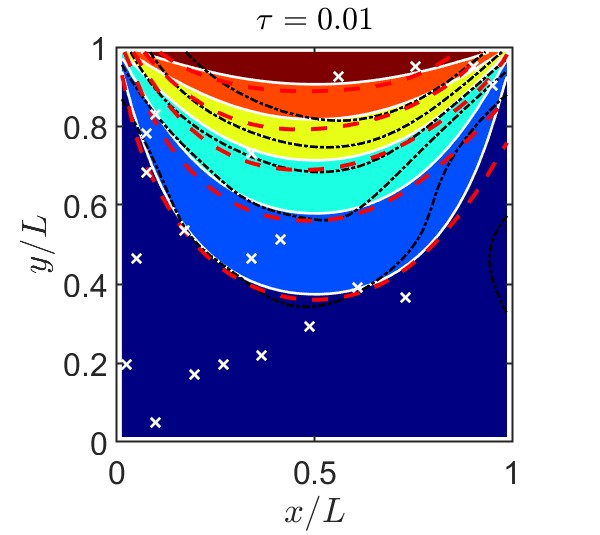}\label{fig:2a}}
\subfigure[]{\includegraphics[width=0.24\textwidth]{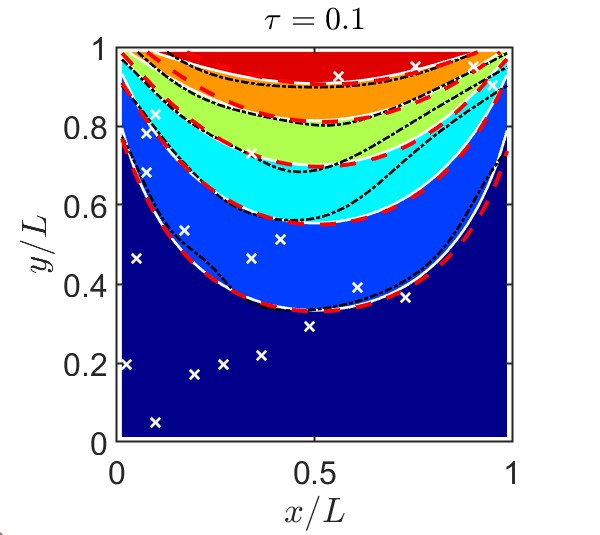}\label{fig:2b}}
\subfigure[]{\includegraphics[width=0.24\textwidth]{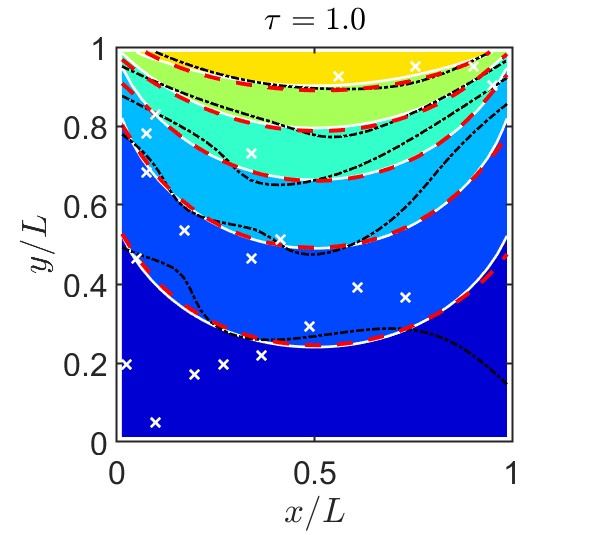}\label{fig:2c}}
\subfigure[]{\includegraphics[width=0.235\textwidth]{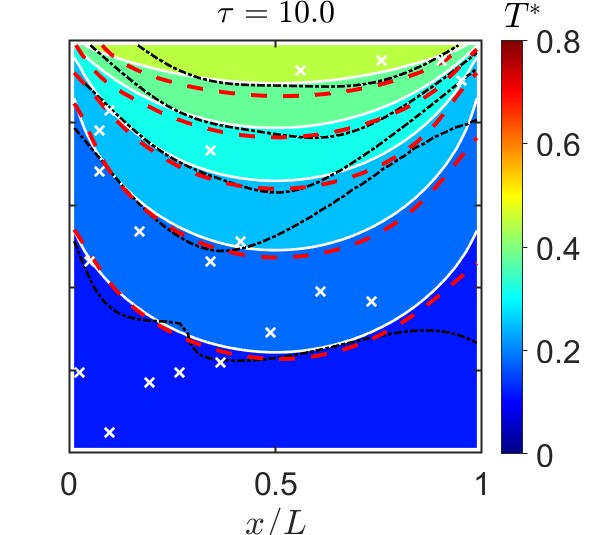}\label{fig:2d}}
\subfigure[]{\includegraphics[width=0.24\textwidth]{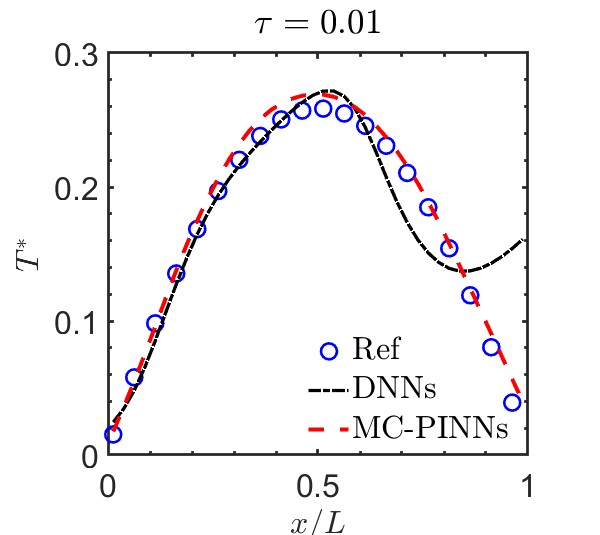}\label{fig:2e}}
\subfigure[]{\includegraphics[width=0.24\textwidth]{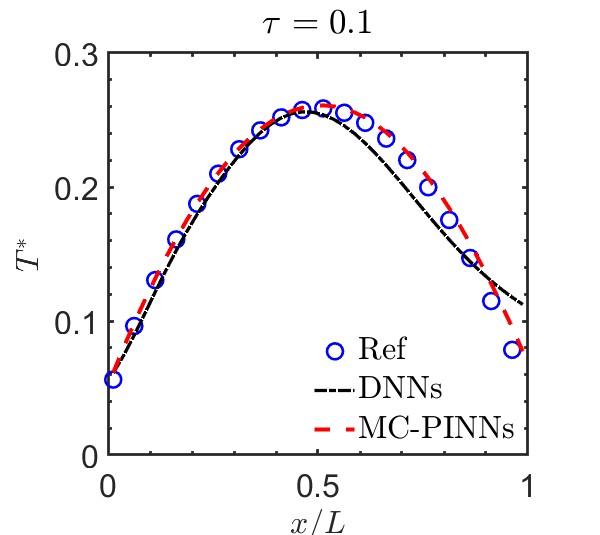}\label{fig:2f}}
\subfigure[]{\includegraphics[width=0.24\textwidth]{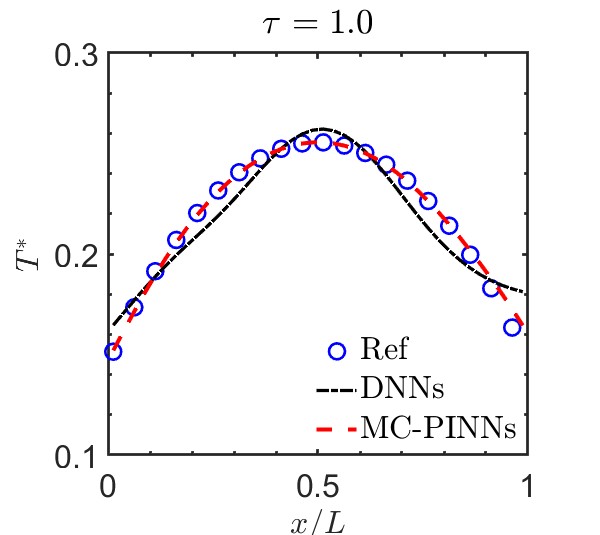}\label{fig:2g}}
\subfigure[]{\includegraphics[width=0.24\textwidth]{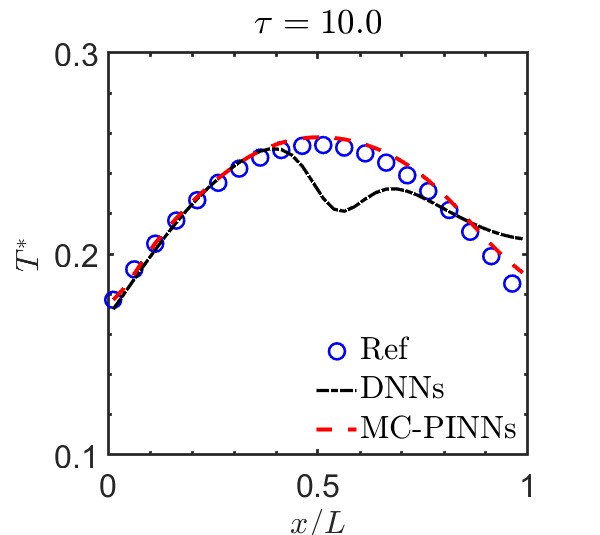}\label{fig:2h}}
\caption{
Quasi-2D multiscale heat conduction: reconstructed temperature fields (top row) and centerline temperature profiles at $y=0.5$ (bottom row) obtained by DNNs (black dashed-dotted line) and MC-PINNs (red dashed line) for different relaxation times: (a, e) $\tau = 0.01$, (b, f) $\tau=0.1$, (c, g) $\tau=1.0$, and (d, h) $\tau=10.0$. Colored background with white solid lines in (a--d): reference solutions~\cite{zhang2023acceleration}; white crosses: sensor locations.}
\label{fig:2}
\end{figure}

Figure~\ref{fig:2} compares the reconstructed temperature fields and centerline profiles at $y=0.5$ obtained by the two methods.
Both the contour plots and the centerline profiles show that MC-PINNs accurately recover the temperature distributions across the full range of relaxation times, whereas the DNN reconstructions exhibit noticeable errors near the boundaries and in regions with sparse observational coverage.
The centerline profiles highlight this difference clearly: MC-PINNs faithfully reproduce the expected heat conduction behavior, while DNNs produce unphysical deviations in poorly observed regions.

To quantify the impact of data sparsity, we evaluate the reconstruction error across varying numbers of sensor points ($N_{\mathrm{sen}}=10,\,20,\,40,$ and $80$) and relaxation times.
For each configuration, five independent runs are performed with randomly regenerated sensor locations, so that both the mean accuracy and the sensitivity to sensor placement can be assessed.

\begin{figure}[htbp]
    \centering
    \includegraphics[width=0.5\textwidth]{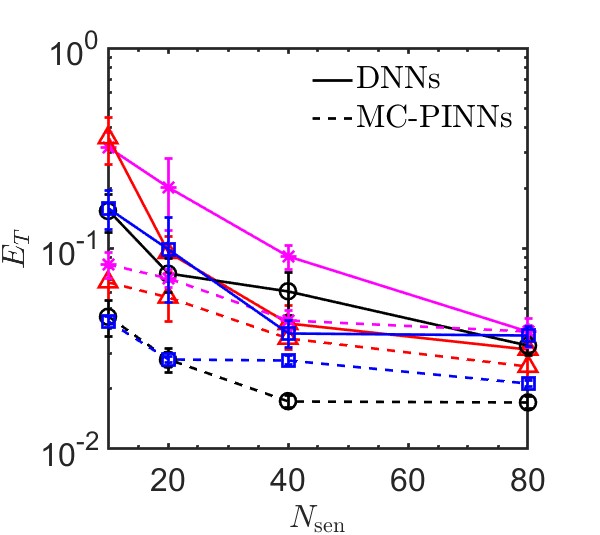}
    \caption{Quasi-2D heat conduction with spatially uniform $\tau$: relative $L_2$ errors of the reconstructed temperature field obtained by DNNs (solid lines) and MC-PINNs (dashed lines) with different numbers of sensor points. Magenta: $\tau=0.01$; black: $\tau=0.1$; blue: $\tau=1.0$; red: $\tau=10.0$. Error bars denote the standard deviation over 5 independent runs.}
    \label{fig:3}
\end{figure}

\begin{table}[htbp]
\centering
\caption{Mean relative $L_2$ error $E_T$ (mean $\pm$ standard deviation) of DNNs and MC-PINNs with different numbers of sensor points.}
\label{tab:ET_comparison}
\setlength{\tabcolsep}{8pt}
\renewcommand{\arraystretch}{1.25}
\begin{tabular}{llcccc}
\toprule[1.5pt]
& & \multicolumn{4}{c}{$E_T$} \\
\cmidrule(lr){3-6}
$\tau$ & Method & $N_{\mathrm{sen}}=10$ & $N_{\mathrm{sen}}=20$ & $N_{\mathrm{sen}}=40$ & $N_{\mathrm{sen}}=80$ \\
\toprule[1.5pt]
\multirow{2}{*}{$0.01$}
  & DNNs      & $\mathbf{0.3208 \pm 0.0587}$ & $0.2022 \pm 0.0790$ & $0.0910 \pm 0.0120$ & $0.0384 \pm 0.0030$ \\
  & MC-PINNs  & $\mathbf{0.0832 \pm 0.0123}$ & $0.0710 \pm 0.0073$ & $0.0440 \pm 0.0048$ & $0.0386 \pm 0.0063$ \\
\midrule[1pt]
\multirow{2}{*}{$0.1$}
  & DNNs      & $\mathbf{0.1539 \pm 0.0331}$ & $0.0749 \pm 0.0148$ & $0.0612 \pm 0.0149$ & $0.0327 \pm 0.0035$ \\
  & MC-PINNs  & $\mathbf{0.0457 \pm 0.0095}$ & $0.0279 \pm 0.0037$ & $0.0173 \pm 0.0012$ & $0.0171 \pm 0.0012$ \\
\midrule[1pt]
\multirow{2}{*}{$1.0$}
  & DNNs      & $\mathbf{0.1592 \pm 0.0353}$ & $0.0984 \pm 0.0446$ & $0.0379 \pm 0.0057$ & $0.0366 \pm 0.0043$ \\
  & MC-PINNs  & $\mathbf{0.0431 \pm 0.0028}$ & $0.0280 \pm 0.0010$ & $0.0277 \pm 0.0014$ & $0.0212 \pm 0.0007$ \\
\midrule[1pt]
\multirow{2}{*}{$10.0$}
  & DNNs      & $\mathbf{0.3564 \pm 0.0936}$ & $0.0954 \pm 0.0194$ & $0.0422 \pm 0.0101$ & $0.0314 \pm 0.0012$ \\
  & MC-PINNs  & $\mathbf{0.0681 \pm 0.0131}$ & $0.0571 \pm 0.0138$ & $0.0356 \pm 0.0043$ & $0.0258 \pm 0.0033$ \\
\bottomrule[1.5pt]
\end{tabular}
\end{table}

Figure~\ref{fig:3} and Table~\ref{tab:ET_comparison} present the relative $L_2$ errors for both methods as functions of the number of sensor points.
While the errors of both methods decrease with increasing sensor density, MC-PINNs consistently yield lower mean errors than DNNs across nearly all tested configurations.
The advantage is most pronounced in the sparse-data regime ($N_{\mathrm{sen}}=10$ or $20$), where DNN errors often reach or exceed $10\%$, whereas MC-PINNs maintain errors at approximately $5\%$ or below.
Moreover, the error bars of DNNs are substantially wider, indicating a strong dependence on the specific placement of sensors.
When observations are sparse or unevenly distributed, DNNs produce large and unstable errors.
In contrast, the narrow error bars of MC-PINNs indicate robust performance and reduced sensitivity to sensor placement. This robustness is a direct consequence of the physical constraints that regularize the solution in unobserved regions.

We further investigate the influence of the number of angular sampling points $B_{\boldsymbol{s}}$ on reconstruction accuracy.
Four relaxation times ($\tau = 0.01,\,0.1,\,1.0$, and $10.0$) are considered, with $B_{\boldsymbol{s}}$ varied as $8,\,16,\,32,\,64,\,128,$ and $256$.
The number of spatial points is fixed at $B_{\boldsymbol{x}}=100$, and $N_{\mathrm{sen}}=20$ sensor points are used.

\begin{figure}[htbp]
    \centering
    \includegraphics[width=0.5\textwidth]{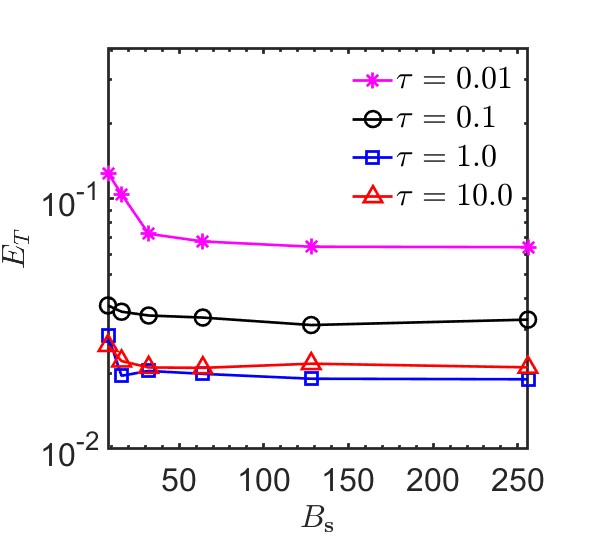}
    \caption{Quasi-2D heat conduction with spatially uniform $\tau$: relative $L_2$ errors of the reconstructed temperature field for MC-PINNs with different numbers of angular sampling points. Magenta: $\tau=0.01$; black: $\tau=0.1$; blue: $\tau=1.0$; red: $\tau=10.0$.}
    \label{fig:2D_Q}
\end{figure}

As shown in Fig.~\ref{fig:2D_Q}, the testing error from MC-PINNs generally decreases with increasing $B_{\boldsymbol{s}}$ for all values of $\tau$.
The improvement saturates once $B_{\boldsymbol{s}}\ge 32$, indicating that relatively sparse angular sampling is sufficient to capture the essential directional information in the inverse reconstruction.
For $\tau = 0.1,\,1.0$, and $10.0$, the relative $L_2$ errors fall below $5\%$ with $B_{\boldsymbol{s}}\ge 32$.
The case $\tau=0.01$ (near-diffusive) shows slightly higher errors, which is expected since the diffusive regime is characterized by near-isotropic angular distributions that require somewhat finer angular resolution. Note that similar results have also been reported in \cite{lin2025monte}.

\subsubsection{Spatially varying relaxation time}

We next consider a quasi-2D case in which the relaxation time varies spatially, to evaluate the robustness of MC-PINNs for a governing equation with heterogeneous thermophysical parameters.
The relaxation time varies only in the $x$-direction according to
\begin{equation}
    \tau(x,y) = \frac{\tau_{\max}+\tau_{\min}}{2} - \frac{\tau_{\max}-\tau_{\min}}{2}\tanh\left({\frac{x-x_c}{2d}}\right),
\end{equation}
with $\tau_{\max}=1.0$, $\tau_{\min}=0.1$, $x_c=L/2$, and $d=0.01$, corresponding to a sharp transition near the domain center.
The prescribed range of $\tau$ lies in the transitional regime, where the thermal field is highly sensitive to spatial variations in the relaxation time.
In the computations, $B_{\boldsymbol{x}}=200$ spatial points and $B_{\boldsymbol{s}}=100$ angular points are used, with $N_{\mathrm{sen}}=20$ sensor measurements for both methods.
The effect of sensor density is also evaluated with $N_{\mathrm{sen}}=10,\,20,\,40$, and $80$, using five independent runs per configuration.

\begin{figure}[H]
\centering
\subfigure[]{\includegraphics[width=0.32\textwidth]{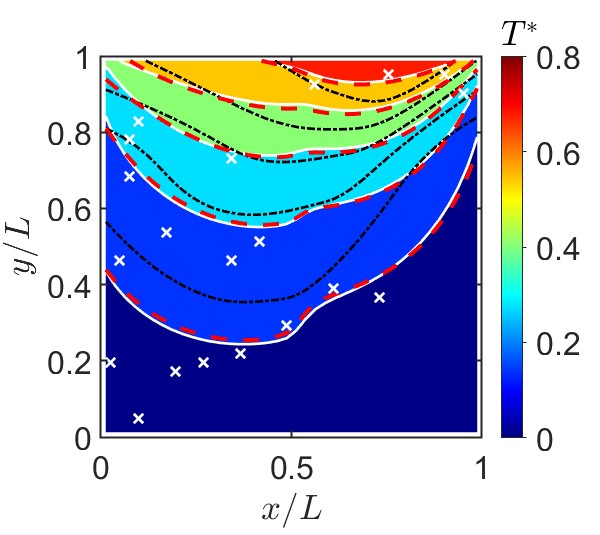}\label{fig:4a}}
\subfigure[]{\includegraphics[width=0.32\textwidth]{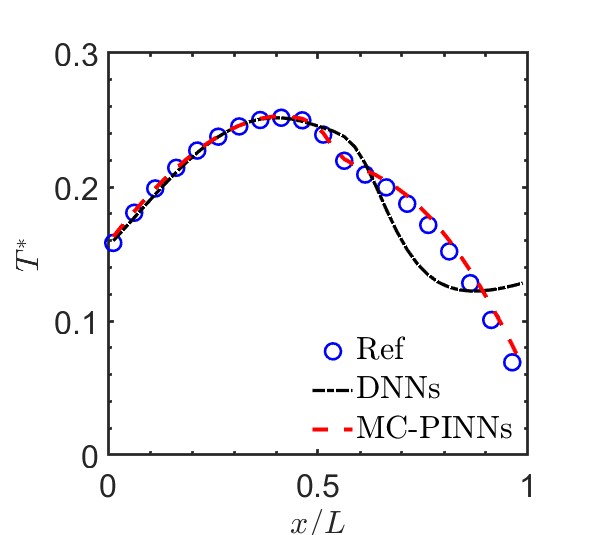}\label{fig:4b}}
\subfigure[]{\includegraphics[width=0.32\textwidth]{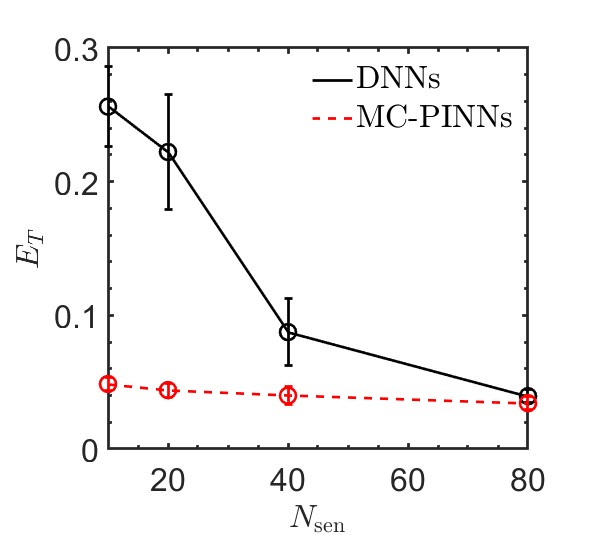}\label{fig:4c}}
\caption{
Quasi-2D heat conduction with spatially varying $\tau$: (a) reconstructed temperature fields and (b) centerline temperature profiles at $y=0.5$ obtained by DNNs (black dash-dotted) and MC-PINNs (red dashed). (c) Relative $L_2$ error $E_T$ of MC-PINNs (dashed) and DNNs (solid) with different numbers of sensor points. Colored background with white solid lines in (a): reference solution~\cite{zhang2023acceleration}; white crosses: sensor locations. Error bars: standard deviation over 5 independent runs.}
\label{fig:4}
\end{figure}

Figures~\ref{fig:4a} and \ref{fig:4b} display the reconstructed temperature fields and centerline profiles at $y=0.5$, respectively.
The sharp variation in $\tau$ near $x=0.5$ induces a pronounced change in the temperature gradient.
MC-PINNs accurately capture this local feature, whereas the purely data-driven DNNs produce an overly smooth solution that fails to resolve the underlying physics.
The relative $L_2$ error comparisons in Fig.~\ref{fig:4c} further quantify this advantage: across all tested sensor configurations, MC-PINNs achieve lower reconstruction errors with significantly smaller variance.

\subsubsection{3D heat conduction in FinFET structures}

To assess the capability of MC-PINNs for realistic device-scale problems, we consider 3D steady-state heat conduction in a bulk FinFET structure, following the configuration in Ref.~\cite{zhang2025effects}.
All geometric dimensions are nondimensionalized using the nanometer as the reference length scale.

\begin{figure}
\centering
\subfigure[]{\includegraphics[width=0.38\textwidth]{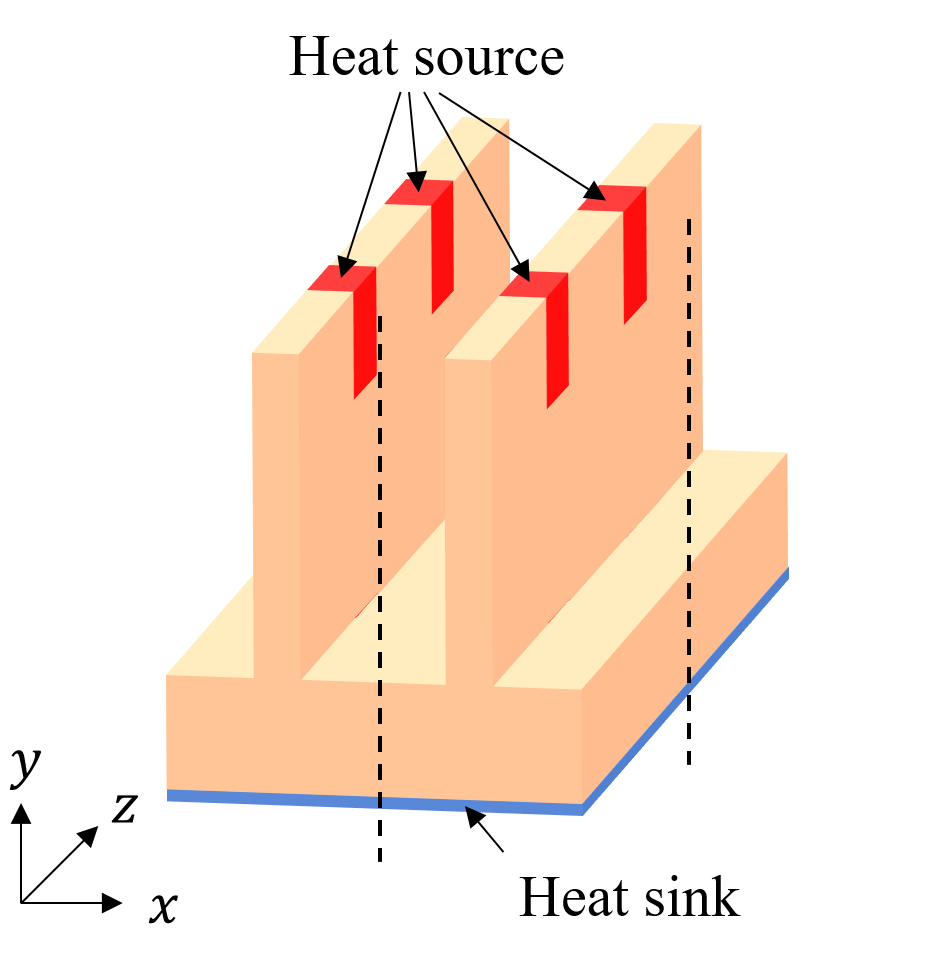}\label{fig:3D1}}
\subfigure[]{\includegraphics[width=0.5\textwidth]{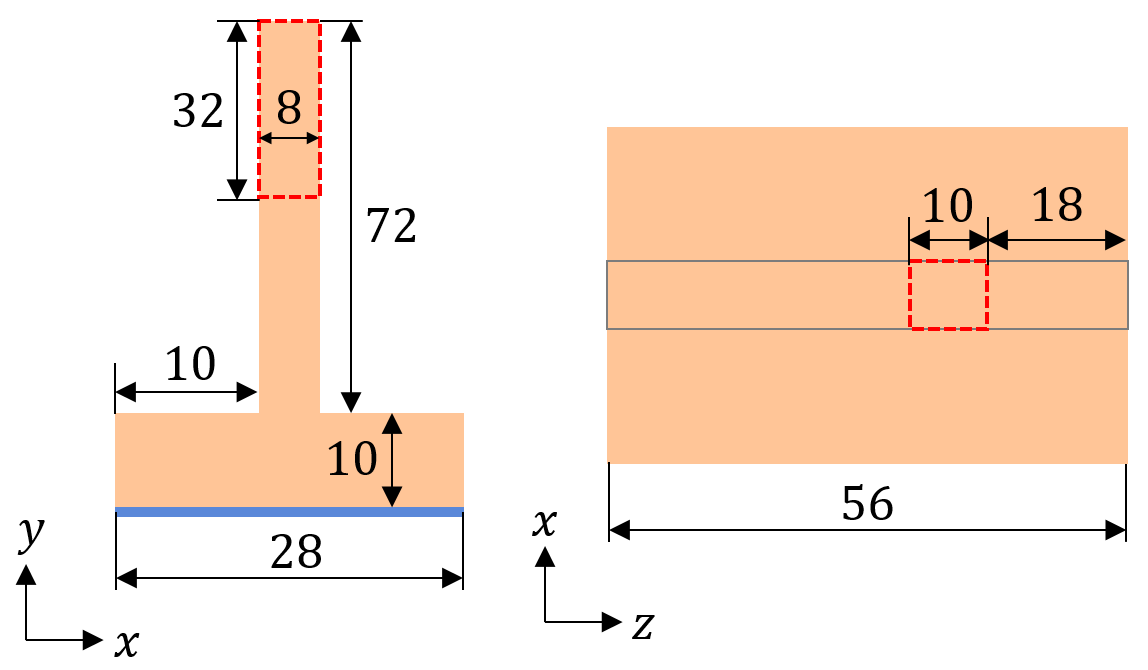}\label{fig:3D2}}
\caption{
Schematic of the bulk FinFET structure and computational domain.
(a) Three-dimensional view of four bulk FinFETs, with gray regions representing SiO$_2$ and light orange regions indicating Si.
(b) Cross-sectional views of the computational domain in the $x$--$y$ and $x$--$z$ planes. Red dashed boxes indicate the heat-source regions.}
\label{fig:3D_schematic}
\end{figure}

The structure, illustrated in Fig.~\ref{fig:3D_schematic}, comprises four bulk FinFETs with symmetrically arranged heat sources embedded in the fin regions.
Exploiting geometric symmetry, only one quarter of the full structure is used as the computational domain.
The bottom surface is held at a fixed temperature $T_w=1$, while all other surfaces are treated as adiabatic and diffusely reflecting boundaries.
The volumetric heat generation is prescribed as
\begin{equation}
\dot{S} =
\begin{cases}
0.005, & \text{if } x \in [10, 18],\, y \in [50, 82],\, z \in [28, 38], \\
0,     & \text{otherwise}.
\end{cases}
\end{equation}
Two cases with different relaxation times, $\tau=10.0$ and $\tau=100.0$, are considered, with the corresponding Knudsen numbers lying in the near-diffusive to transitional regimes (using $L_c=82$, the maximum device dimension along the primary heat-conduction path). In each test case, we assume $N_{\mathrm{sen}}=200$ sensors are randomly placed in the device to provide in situ measurements of the temperature.

\begin{figure}[htbp]
\centering
\subfigure[]{\includegraphics[width=0.37\textwidth]{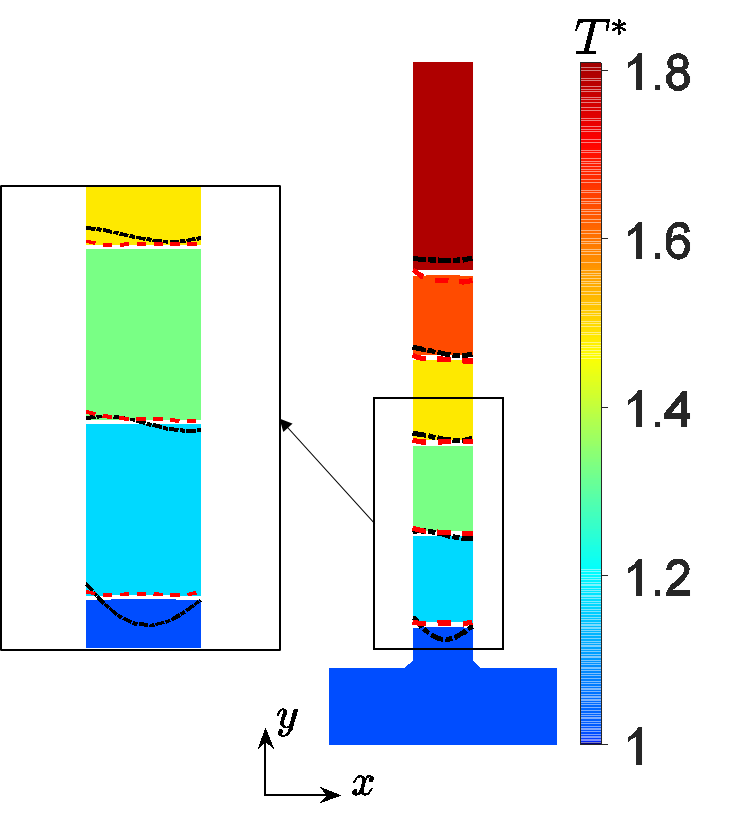}}
\subfigure[]{\includegraphics[width=0.5\textwidth]{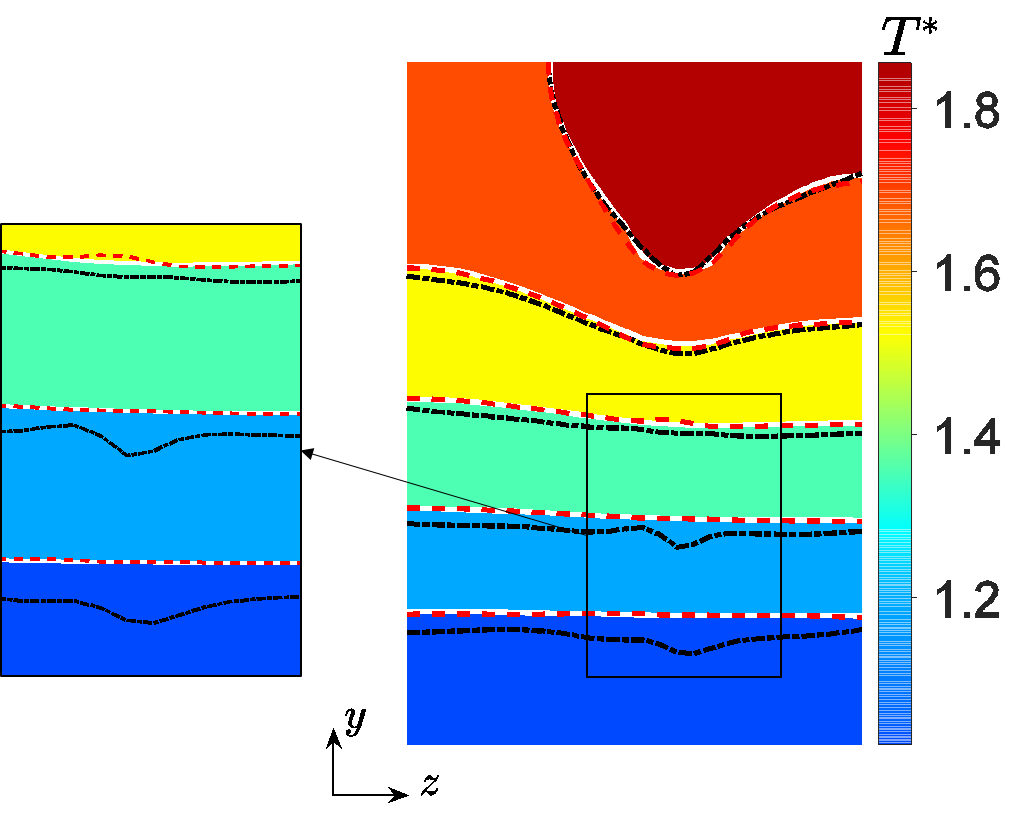}}
\subfigure[]{\includegraphics[width=0.32\textwidth]{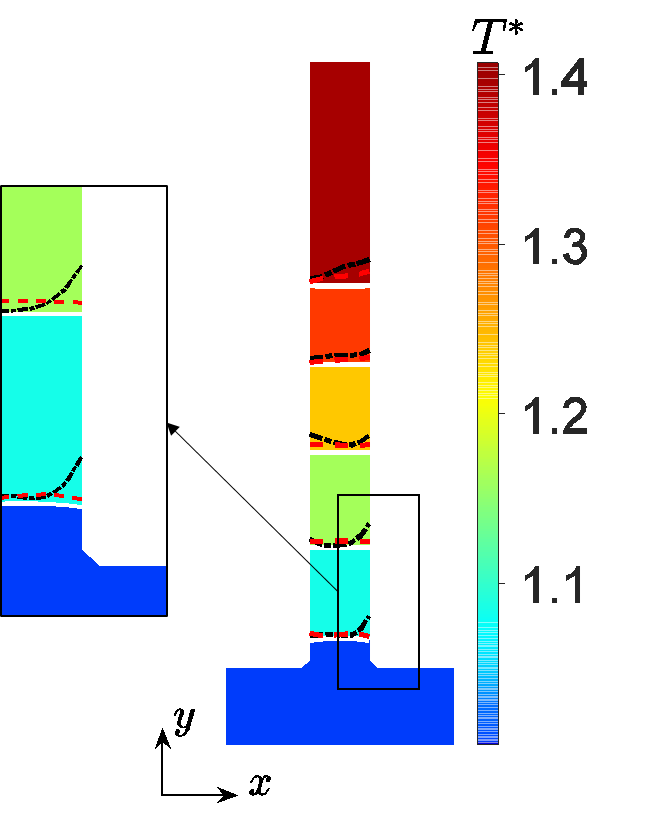}}
\subfigure[]{\includegraphics[width=0.5\textwidth]{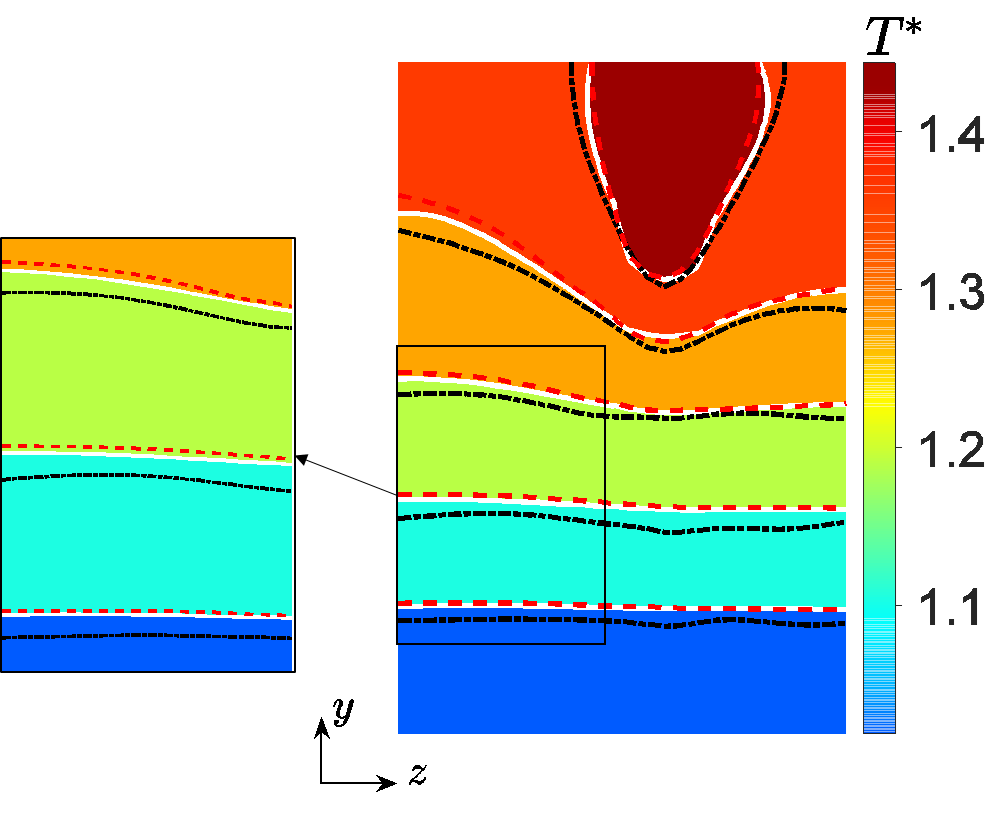}}
\caption{
3D heat conduction in a FinFET structure: reconstructed temperature slices in the $x$--$y$ plane at $z=29$ and the $y$--$z$ plane at $x=14.5$ for (a, b) $\tau = 10.0$ and (c, d) $\tau = 100.0$.
Colored background with white solid lines: reference solutions~\cite{zhang2025effects}; black dotted-dashed: DNNs; red dashed: MC-PINNs.}
\label{fig:5}
\end{figure}

For the MC-PINN computations, $B_{\boldsymbol{x}}=500$ spatial points and $B_{\boldsymbol{s}}=100$ angular points are randomly sampled at each iteration, with $N_{\mathrm{sen}}=200$ sensor points used for training.
The reconstructed 3D thermal field is visualized through temperature slices in the $x$--$y$ plane at $z=29$ and in the $y$--$z$ plane at $x=14.5$. As shown in Fig.~\ref{fig:5} and Table~\ref{tab:tab1}, MC-PINNs consistently achieve closer agreement with the reference solutions than DNNs for both values of $\tau$, with relative $L_2$ errors approximately 2-3 times smaller.
These results confirm that the proposed framework scales effectively to three-dimensional, device-scale geometries.

\begin{table}[htbp]
\caption{
3D FinFET heat conduction: relative $L_2$ errors of the temperature field reconstructed by DNNs and MC-PINNs.}
\label{tab:tab1}
\newcolumntype{C}{>{\centering\arraybackslash}X}
\begin{tabularx}
{\textwidth}{CCC}
\toprule
  & $\tau=10.0$ & $\tau=100.0$\\
\midrule
$E_T(\mathrm{DNNs})$ & 0.948\% & 0.807\%\\
\midrule
$E_T(\mathrm{MC\mbox{-}PINNs})$ & 0.380\% & 0.309\%\\
\bottomrule
\end{tabularx}
\end{table}

Several observations merit further discussion.
First, examining the temperature slices in Fig.~\ref{fig:5}, both MC-PINNs and DNNs capture the overall thermal structure, including the elevated temperature in the fin regions where the heat sources are located and the gradual temperature decay toward the isothermal bottom surface.
However, the DNN reconstructions exhibit visible discrepancies in the $y$--$z$ plane (Figs.~\ref{fig:5}b and \ref{fig:5}d), particularly in the lower portion of the domain near the isothermal boundary, where the temperature contours predicted by DNNs deviate from the reference isotherms.
In contrast, the MC-PINNs contours closely follow the reference isotherms throughout the domain, including in regions far from sensor locations.
This behavior is consistent with the observations in the quasi-2D cases: purely data-driven models struggle to extrapolate accurately beyond the vicinity of the sensors, whereas the embedded BTE residual in MC-PINNs propagates information along characteristic directions and regularizes the solution in unobserved regions.

Second, both methods achieve lower relative errors for $\tau=100.0$ than for $\tau=10.0$ (Table~\ref{tab:tab1}).
Using the characteristic length $L_c=82$ and the relation $\mathrm{Kn}=v_g\tau/L_c$ with $v_g=1$, the corresponding Knudsen numbers are $\mathrm{Kn}\approx0.12$ for $\tau=10.0$ and $\mathrm{Kn}\approx1.22$ for $\tau=100.0$.
Both cases lie within the near-diffusive to transitional range, but their transport characteristics differ in ways that affect reconstructability.
For $\tau=10.0$ ($\mathrm{Kn}\approx0.12$, near-diffusive), phonon scattering is frequent and heat conduction is predominantly local, following an approximately Fourier-like behavior.
The temperature field is characterized by relatively sharp thermal gradients confined near the embedded heat sources in the fin regions.
For $\tau=100.0$ ($\mathrm{Kn}\approx1.22$, transitional), the phonon mean free path ($\lambda=100$) is comparable to the device dimensions, and non-local transport effects become significant.
Phonons emitted from the heat sources travel further before thermalizing, distributing thermal energy more broadly across the domain. The resulting macroscopic temperature field exhibits less localized gradients and is therefore easier to reconstruct from sparse measurements for both methods.
A closer examination of the temperature slices in Fig.~\ref{fig:5} reveals that the DNN predictions exhibit noticeable discrepancies in two regions in particular.
The first is near the heat sources (the fin regions in the $x$--$y$ plane at $z=29$), where the DNN contours deviate from the reference isotherms around the high-temperature zones, indicating that the purely data-driven model fails to adequately resolve the localized thermal gradients in the vicinity of the sources.
The second is near the isothermal bottom boundary (visible in the $y$--$z$ plane at $x=14.5$), where the DNN-predicted temperature contours drift away from the reference solution in the lower portion of the domain.
Both types of discrepancy are more pronounced for $\tau=100.0$ (Figs.~\ref{fig:5}c and \ref{fig:5}d): in the transitional regime, the DNN struggles to extrapolate from interior sensor measurements to the unobserved boundary, and the reconstructed temperature near the bottom surface deviates visibly from the prescribed isothermal condition.
In contrast, the MC-PINNs reconstructions remain in close agreement with the reference solution in both the near-source and near-boundary regions across both values of $\tau$.
This can be attributed to the BTE residual in the loss function, which enforces the governing transport physics throughout the domain and propagates information from the interior sensors toward the boundaries along the characteristic directions of the kinetic equation, compensating for the absence of boundary data in Problem~(1).

Third, it is instructive to consider the ratio of sensor points to the problem dimensionality in this 3D case.
With $N_{\mathrm{sen}}=200$ sensor points distributed in a 3D domain of approximately $82\times 82\times 40$ (in dimensionless units), the effective sensor density is substantially lower than in the quasi-2D cases considered earlier.
For the quasi-2D case with $N_{\mathrm{sen}}=20$ in a $1\times 1$ domain, the linear sensor density is approximately $4.5$ per unit length.
In the FinFET case, with $N_{\mathrm{sen}}=200$ distributed in a domain of characteristic dimension $L_c=82$ along the primary heat-conduction direction, the corresponding density is only about $1.4$ per unit length, roughly three times sparser.
Despite this reduced effective sensor density, MC-PINNs achieve relative errors below $0.4\%$, while DNNs, which lack physical constraints, incur errors near $1\%$.
This comparison shows that embedding physical knowledge is more beneficial when the ratio of observational data to problem size is small, a regime of considerable practical relevance where measurements are inherently limited.

Fourth, the FinFET geometry introduces geometric complexity absent in the quasi-2D square domain: multiple material interfaces (Si and SiO$_2$), embedded heat sources of finite extent, and non-rectangular boundaries.
In grid-based PINN formulations for the BTE, such geometric complexity would necessitate careful mesh generation and potentially local grid refinement near interfaces and small features.
The MC-PINNs framework circumvents these difficulties through its mesh-free sampling strategy: collocation points are simply drawn from the computational domain without regard to geometric features, and the two-step sampling naturally handles the higher-dimensional temporal-spatial-angular space.
This demonstrates a practical advantage of the Monte Carlo approach for device-scale thermal analysis, where geometric complexity is the norm rather than the exception.

Finally, we note that the 200 sensor points used here represent only a modest fraction of the total degrees of freedom in the problem.
The reference solution, obtained using the discrete ordinates method with $48\times 48$ angular quadrature points and a spatial resolution of $dx=dy=1$ and $dz=2$, involves on the order of $10^5$ grid points, each with angular degrees of freedom.
That MC-PINNs can reconstruct the thermal field to within $0.4\%$ accuracy using measurements at only 200 spatial locations, without knowledge of the boundary conditions, is due to the strong regularization provided by the BTE residual.
The governing equation constrains the admissible solution to a low-dimensional manifold embedded in the high-dimensional discretized function space, and the sensor measurements serve primarily to select the correct point on this manifold.
This property is general to physics-informed inverse methods and is not specific to the FinFET configuration, but the 3D device-scale example illustrates its practical value.

\subsection{Simultaneous inference of thermal field and unknown relaxation time}\label{sec:results_unknown_tau}

We now turn to Problem (2), where the boundary conditions are prescribed but the relaxation time $\tau$ is unknown and must be inferred jointly with the thermal field.
This setting is adopted to reduce the degree of ill-posedness: if both the boundary conditions and $\tau$ were unknown, sparse temperature measurements alone would generally be insufficient to uniquely determine all unknowns.

\subsubsection{Inferring spatially uniform $\tau$ as a trainable parameter}

We first examine the case where $\tau$ is spatially uniform but unknown, treated as a trainable parameter optimized jointly with the network weights.
Steady-state heat conduction in a quasi-1D film of length $L=1$ is considered, with ground-truth relaxation times $\tau=0.1,\,1.0$, and $10.0$.
The temperatures at the left and right boundaries ($x=0$ and $x=L$) are fixed at $T_h=1$ and $T_c=0$, respectively, and the thermalization boundary condition is imposed.
For each case, $N_{\mathrm{sen}}=2$ sensor points, $B_{\boldsymbol{x}}=40$ interior collocation points, and $B_{\boldsymbol{s}}=64$ angular samples are used.

\begin{figure}[H]
\centering
\subfigure[]{\includegraphics[width=0.41\textwidth]{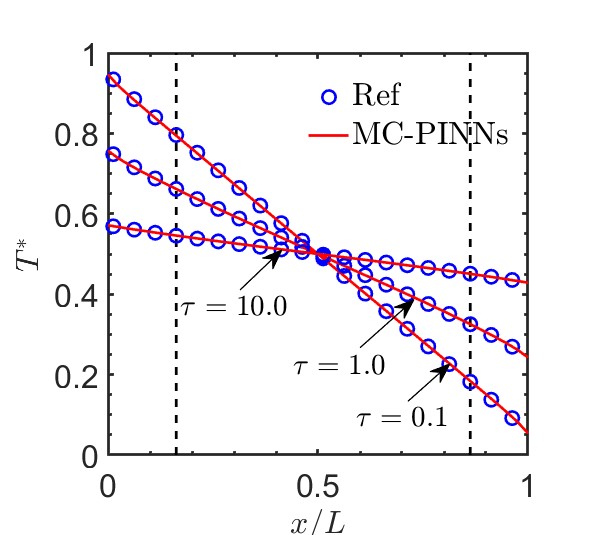}\label{fig:6a}}
\subfigure[]{\includegraphics[width=0.4\textwidth]{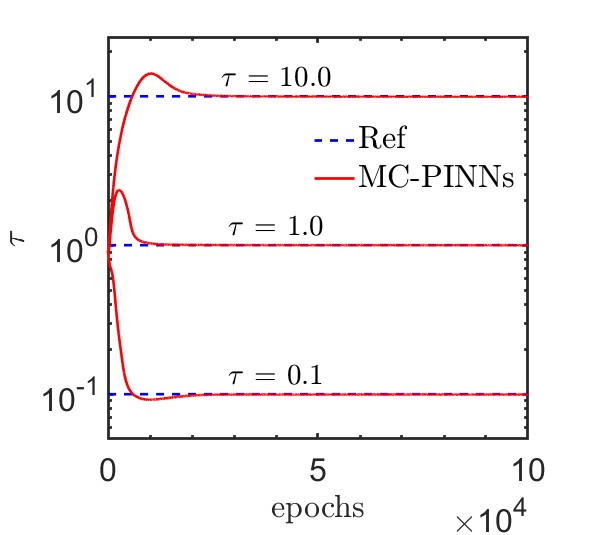}\label{fig:6b}}
\caption{Quasi-1D heat conduction with unknown spatially uniform $\tau$: (a) reconstructed temperature fields by DNNs (black dashed-dotted) and MC-PINNs (red solid) for different ground-truth $\tau$. Black vertical dashed lines indicate sensor locations. (b) Inferred $\tau$ values compared with ground truth. Blue dashed: reference solutions~\cite{zhang2019implicit}; red solid: MC-PINNs prediction.}
\label{fig:6}
\end{figure}

\begin{table}[htbp]
\caption{
Quasi-1D heat conduction with unknown spatially uniform $\tau$: relative $L_2$ errors of the reconstructed temperature and inferred $\tau$ obtained by MC-PINNs.}
\label{tab:tab2}
\newcolumntype{C}{>{\centering\arraybackslash}X}
\begin{tabularx}
{\textwidth}{CCCC}
\toprule
  & $\tau=0.1$ & $\tau=1.0$ & $\tau=10.0$\\
\midrule
$E_T$      & 0.288\% & 0.079\% & 0.068\%\\
$E_{\tau}$ & 0.500\% & 0.100\% & 0.400\%\\
\bottomrule
\end{tabularx}
\end{table}

Figure~\ref{fig:6a} shows that the temperature profiles reconstructed by MC-PINNs closely match the reference solutions~\cite{zhang2019implicit} for all three cases, despite using only two sensor points.
The predicted distributions correctly capture the boundary-layer effects that produce nonlinear temperature profiles near the walls (most evident at $\tau=1.0$).
Figure~\ref{fig:6b} and Table~\ref{tab:tab2} show that the framework also infers $\tau$ with high accuracy: the relative errors in $\tau$ are below $0.5\%$ for all cases, and the temperature errors remain below $0.3\%$.

We further examine the influence of angular sampling size $B_{\boldsymbol{s}}$ under the same quasi-1D setting, with $N_{\mathrm{sen}}=2$ and $B_{\boldsymbol{x}}=40$ fixed, while $B_{\boldsymbol{s}}$ is varied as $8,\,16,\,32,\,64,\,128$, and $256$.

\begin{figure}[H]
\centering
\subfigure[]{\includegraphics[width=0.4\textwidth]{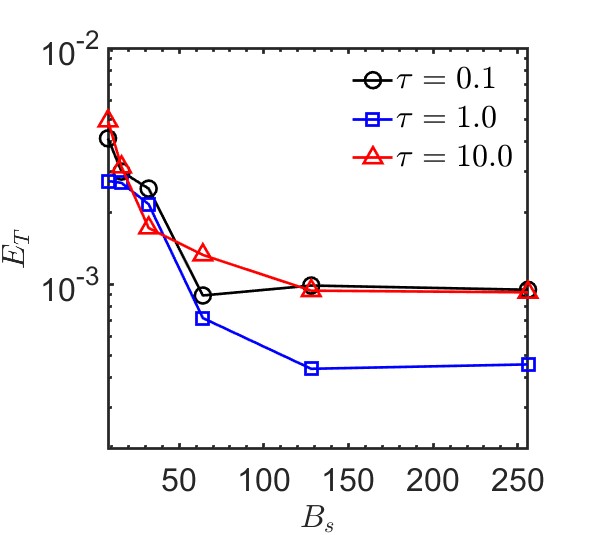}\label{fig:1D_Qa}}
\subfigure[]{\includegraphics[width=0.4\textwidth]{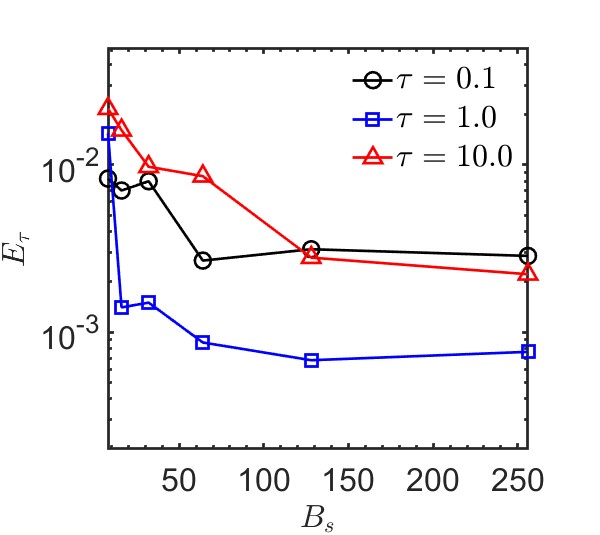}\label{fig:1D_Qb}}
\caption{
Quasi-1D heat conduction with unknown spatially uniform $\tau$: relative $L_2$ errors of (a) the reconstructed temperature and (b) the inferred $\tau$ obtained by MC-PINNs with different numbers of angular sampling points.
Black, blue, and red lines correspond to $\tau=0.1$, $1.0$, and $10.0$, respectively.}
\label{fig:1D_Q}
\end{figure}

As shown in Fig.~\ref{fig:1D_Q}, increasing $B_{\boldsymbol{s}}$ generally improves both temperature reconstruction and $\tau$ inference.
Accurate results are already achieved with relatively modest angular sampling: for $B_{\boldsymbol{s}}\ge 32$, the temperature error remains below $0.5\%$ and the $\tau$ error below $1\%$ in most cases.
This insensitivity to the angular resolution is particularly valuable for inverse problems, where the transport regime is not known a priori.
In grid-based PINNs or conventional deterministic numerical methods, the angular discretization must be chosen according to the expected Knudsen number, but this choice becomes ambiguous when $\tau$ is unknown.
MC-PINNs avoid this difficulty through the mesh-free sampling strategy, which provides a uniform treatment across transport regimes.

A quasi-2D case is also examined with reference $\tau$ values of $0.1$ and $1.0$.
For both cases, $N_{\mathrm{sen}} = 40$ sensor points for the temperature are randomly sampled, with $B_{\boldsymbol{x}} = 100$ interior collocation points and $B_{\boldsymbol{s}}=100$ angular points per training step.
The top boundary is maintained at $T_h = 1$, while the remaining boundaries are set to $T_c = 0$.

\begin{figure}[htbp]
\centering
\subfigure[]{\includegraphics[width=0.32\textwidth]{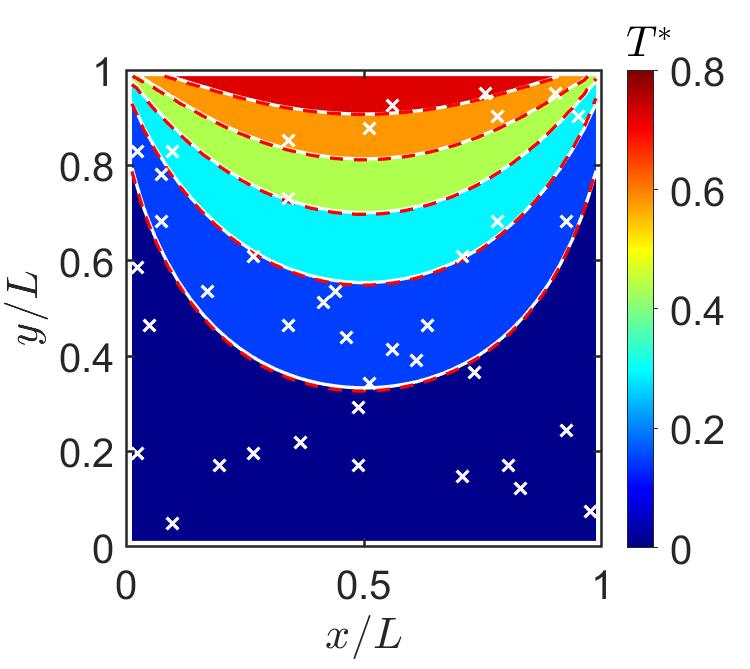}\label{fig:7a}}
\subfigure[]{\includegraphics[width=0.32\textwidth]{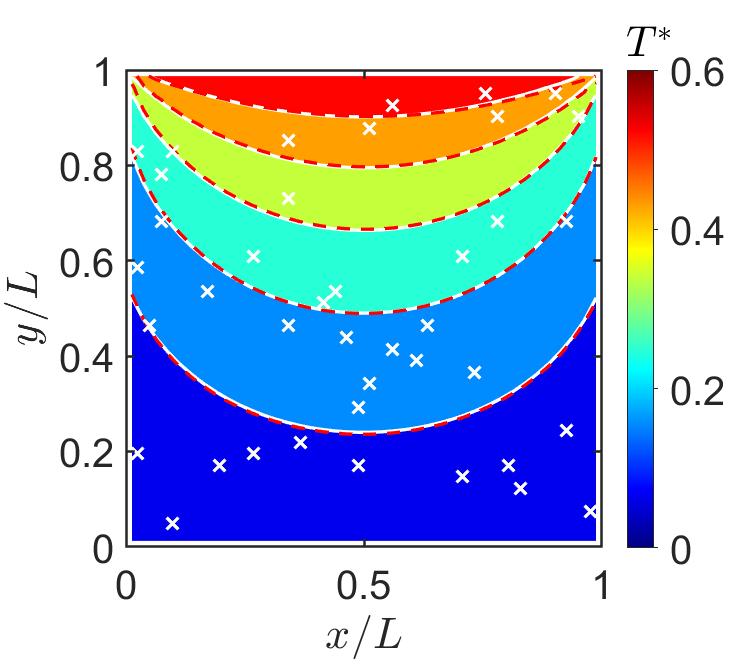}\label{fig:7b}}
\subfigure[]{\includegraphics[width=0.305\textwidth]{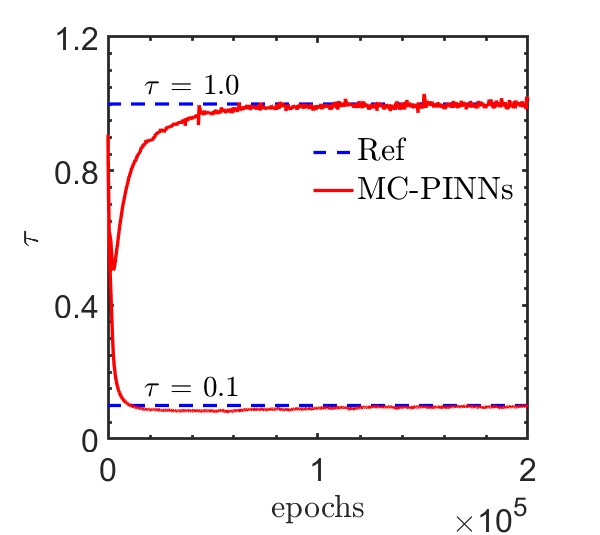}\label{fig:7c}}
\caption{
Quasi-2D heat conduction with unknown spatially uniform $\tau$: reconstructed temperature fields by MC-PINNs (red dashed) for (a) $\tau = 0.1$ and (b) $\tau = 1.0$. (c) Inferred $\tau$ compared with ground truth. Colored background with white solid lines in (a, b): reference solutions~\cite{zhang2023acceleration}; white crosses: sensor locations.}
\label{fig:7}
\end{figure}

\begin{table}[htbp]
\caption{
Quasi-2D heat conduction with unknown spatially uniform $\tau$: relative $L_2$ errors of reconstructed temperature and inferred $\tau$ obtained by MC-PINNs.}
\label{tab:tab3}
\newcolumntype{C}{>{\centering\arraybackslash}X}
\begin{tabularx}
{\textwidth}{CCCC}
\toprule
  & $\tau=0.1$ & $\tau=1.0$\\
\midrule
$E_T$      & 1.090\% & 1.771\%\\
$E_{\tau}$ & 0.891\% & 0.449\%\\
\bottomrule
\end{tabularx}
\end{table}

Figure~\ref{fig:7} and Table~\ref{tab:tab3} confirm that MC-PINNs accurately reconstruct the temperature field from limited sensor data in the quasi-2D setting, even when the relaxation time is unknown.
The inferred $\tau$ values are in close agreement with the ground truth, and the reconstructed solutions are consistent with the reference results.
These findings demonstrate that the embedded physical constraints effectively regularize the inverse problem, enabling simultaneous field reconstruction and parameter identification with sparse measurements.

\subsubsection{Inferring spatially varying $\tau$ using sub-networks}

We now apply the MC-PINNs to the more challenging scenarios where the relaxation time varies spatially and is modeled as a function $\tau(\boldsymbol{x})$ using a sub-network, as described in Sec.~\ref{sec:mcpinns}.
Both quasi-1D and quasi-2D steady-state problems are considered.

\textbf{Quasi-1D case.}
Heat conduction across a film of length $L=1$ is considered, with prescribed boundary temperatures $T_c=0$ at $x=0$ and $T_h=1$ at $x=L$.
The ground-truth $\tau(x)$ varies smoothly according to Eq.~\eqref{eq:tau},
\begin{equation}\label{eq:tau}
    \tau(x) = \frac{\tau_{\max} + \tau_{\min}}{2} - \frac{\tau_{\max} - \tau_{\min}}{2} \tanh \left(\frac{x-x_c}{2d} \right),
\end{equation}
with $\tau_{\min}=0.1$, $\tau_{\max}=1.0$, $x_c = L/2$, and $d = 0.01$.
A total of $N_{\mathrm{sen}} = 40$ sensor points are randomly distributed, with $B_{\boldsymbol{x}} = 100$ spatial points and $B_{\boldsymbol{s}} = 64$ angular points sampled at each training iteration.

\begin{figure}[htbp]
\centering
\subfigure[]{\includegraphics[width=0.4\textwidth]{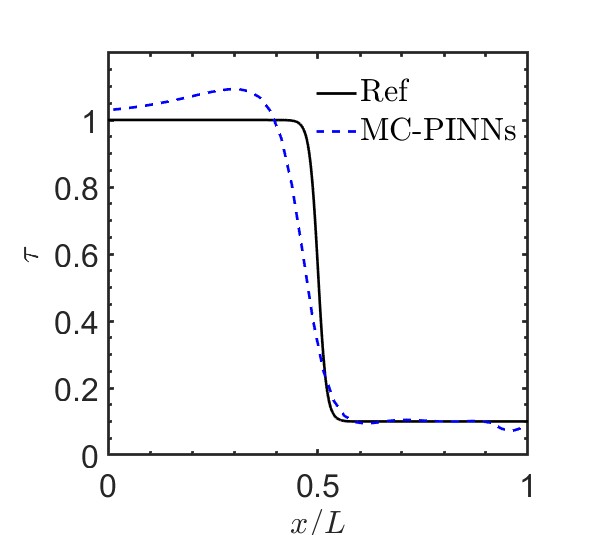}\label{fig:8a}}
\subfigure[]{\includegraphics[width=0.4\textwidth]{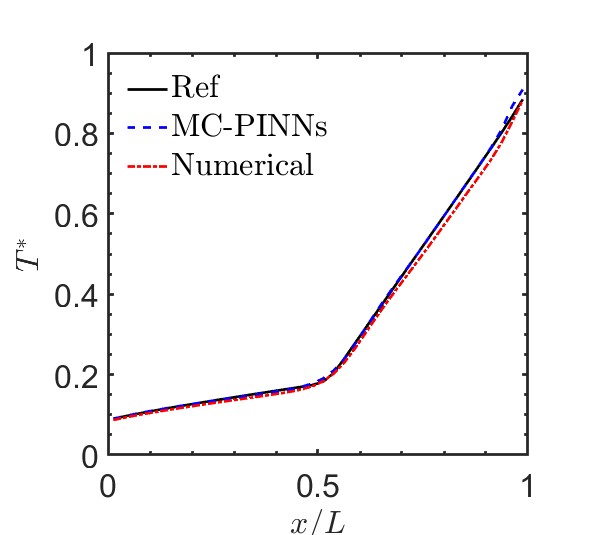}\label{fig:8b}}
\caption{
Quasi-1D heat conduction with unknown spatially varying $\tau$: (a) inferred $\tau(x)$ compared with ground truth, and (b) temperature fields reconstructed by MC-PINNs and recomputed by the numerical solver using the inferred $\tau(x)$.}
\label{fig:8}
\end{figure}

As shown in Fig.~\ref{fig:8a}, the inferred $\tau(x)$ agrees well with the ground truth on the right side of the film ($\tau\approx 0.1$), but exhibits noticeable discrepancies on the left side and near the center, where $\tau$ is larger or varies sharply.
Despite these local deviations, the inferred profile captures the general variation trend and correctly identifies the location of the transition layer.
Figure~\ref{fig:8b} shows that the temperature field reconstructed by MC-PINNs agrees well with the reference solution~\cite{zhang2019implicit}, despite the local errors in $\tau$.
Quantitatively, $E_\tau=44.6\%$, while the directly reconstructed temperature field has an error of $E_T(\mathrm{MC\mbox{-}PINNs})=1.472\%$.
When the inferred $\tau(x)$ is used in the numerical solver, the recomputed temperature field has an error of $E_T(\mathrm{Numerical})=3.284\%$.
These results indicate that accurate pointwise identification of $\tau(x)$ remains challenging, particularly in regions where the macroscopic field is insensitive to variations in $\tau$, while the inferred distribution nonetheless captures the dominant thermal response.

The larger deviations in $\tau(x)$ on the left side and near the transition are attributable to two factors.
First, near the center, the sharp spatial variation in $\tau$ is difficult to resolve from sparse sensor data.
Second, in regions where $\tau$ is large, the scattering term becomes relatively insensitive to changes in $\tau$ (the collision frequency scales as $1/\tau$), so variations in $\tau$ have only a weak influence on the distribution function and the resulting temperature field.
Compared with the spatially uniform case (Sec.~\ref{sec:results_unknown_tau}), the spatially varying $\tau(x)$ introduces a much larger solution space, increasing the possibility of non-unique reconstructions from sparse macroscopic observations.
A detailed sensitivity analysis is provided in~\ref{sec:appendix_a}.

\textbf{Quasi-2D case.}
We further evaluate the sub-network approach in a quasi-2D setting, with reference $\tau = 0.1$ and $\tau = 1.0$ uniform in the ground truth, but treated as unknown spatially varying functions during inference.
The temperature is set to $T_h = 1$ at the top boundary and $T_c = 0$ at the remaining boundaries.
A total of $N_{\mathrm{sen}}=40$ sensor points are randomly distributed, with $B_{\boldsymbol{x}} = 200$ spatial points and $B_{\boldsymbol{s}} = 100$ angular directions per training step.
The losses associated with $q_x$ and $q_y$ in $\mathcal{L}_{\Phi}$ are assigned an additional weight of $w = 10$ due to their relatively small magnitudes compared with temperature.

\begin{figure}[H]
\centering
\subfigure[]{\includegraphics[width=0.24\textwidth]{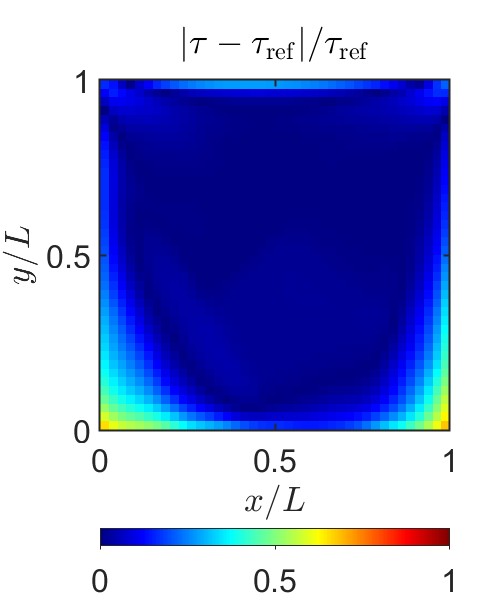}\label{fig:9a}}
\subfigure[]{\includegraphics[width=0.24\textwidth]{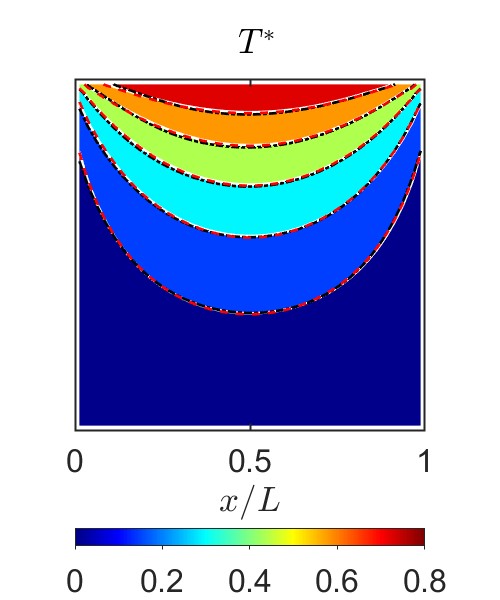}\label{fig:9b}}
\subfigure[]{\includegraphics[width=0.24\textwidth]{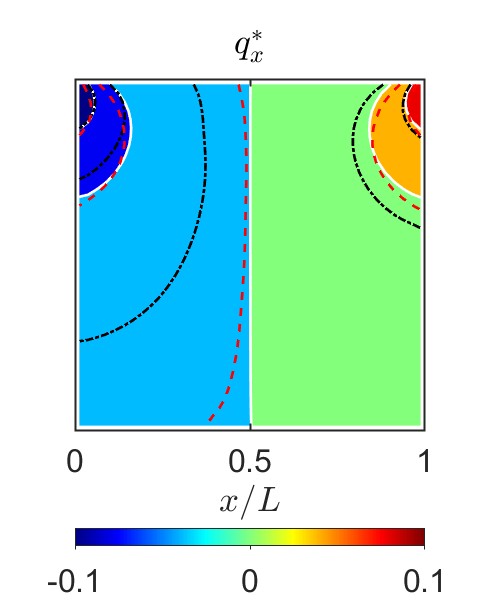}\label{fig:9c}}
\subfigure[]{\includegraphics[width=0.24\textwidth]{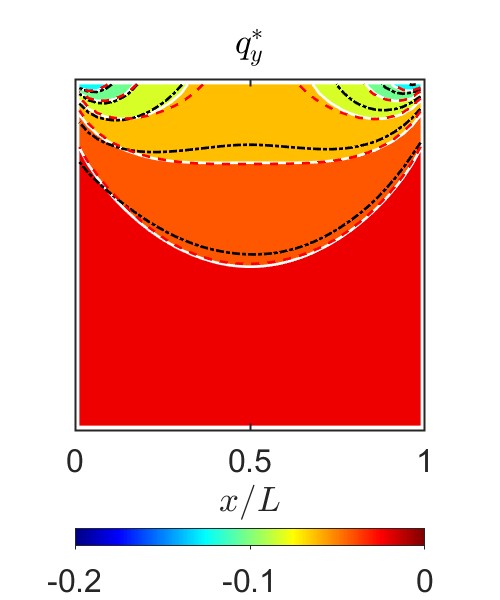}\label{fig:9d}}
\subfigure[]{\includegraphics[width=0.24\textwidth]{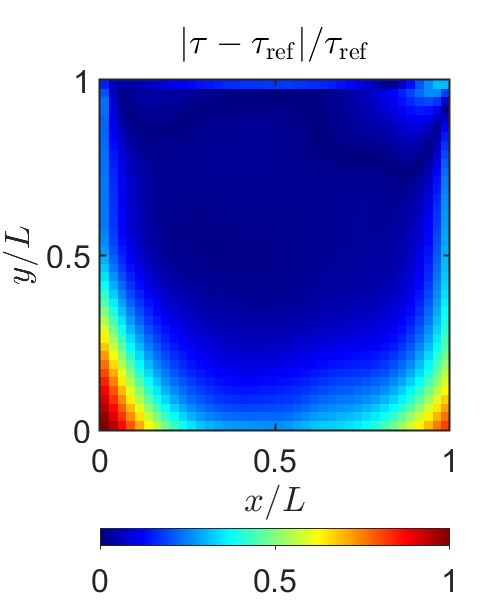}\label{fig:9e}}
\subfigure[]{\includegraphics[width=0.24\textwidth]{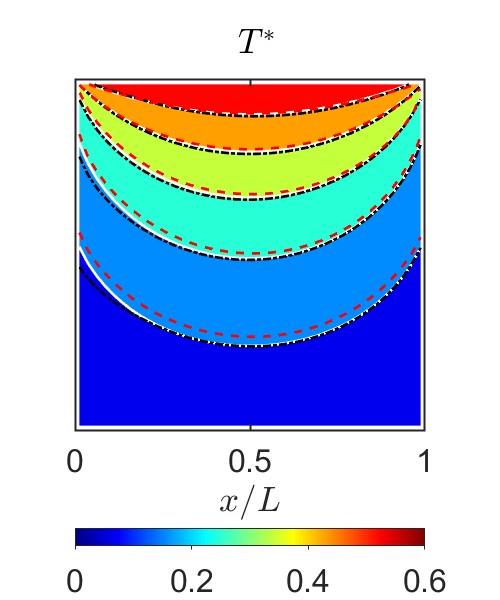}\label{fig:9f}}
\subfigure[]{\includegraphics[width=0.24\textwidth]{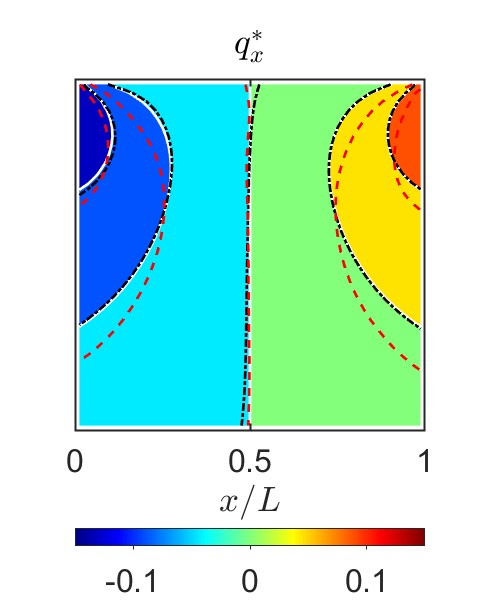}\label{fig:9g}}
\subfigure[]{\includegraphics[width=0.24\textwidth]{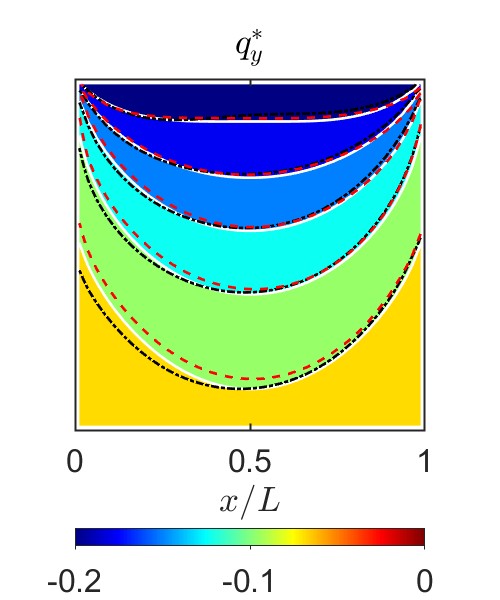}\label{fig:9h}}
\caption{
Quasi-2D heat conduction with unknown spatially varying $\tau$ inferred using sub-networks.
Ground-truth $\tau = 0.1$ (a--d) and $\tau = 1.0$ (e--h).
(a, e) Relative discrepancy between inferred and true $\tau(x,y)$;
(b, f) reconstructed temperature fields from MC-PINNs (red dashed) and numerical results using inferred $\tau(x,y)$ (black dash-dotted);
(c, d, g, h) reconstructed heat fluxes from MC-PINNs (red dashed) and numerical results using inferred $\tau(x,y)$ (black dash-dotted).
Colored background with white solid lines: reference solutions~\cite{zhang2023acceleration}.}
\label{fig:9}
\end{figure}

\begin{table}[htbp]
\caption{
Quasi-2D heat conduction with unknown spatially varying $\tau$: relative $L_2$ errors of inferred $\tau(x,y)$, reconstructed temperature, and heat flux from MC-PINNs.}
\label{tab:tab4}
\newcolumntype{C}{>{\centering\arraybackslash}X}
\begin{tabularx}
{\textwidth}{CCCCC}
\toprule
 & $E_\tau$ & $E_T$ & $E_{q_x}$ & $E_{q_y}$\\
\midrule
$\tau_\mathrm{ref} = 0.1$ & 32.950\% & 0.981\% & 13.582\% & 2.779\%\\
$\tau_\mathrm{ref} = 1.0$ & 20.593\% & 1.963\% & 18.763\% & 1.385\%\\
\bottomrule
\end{tabularx}
\end{table}

\begin{table}[htbp]
\caption{
Quasi-2D heat conduction with unknown spatially varying $\tau$: relative $L_2$ errors of temperature and heat flux recomputed by the numerical solver using the $\tau$ inferred by MC-PINNs.}
\label{tab:tab5}
\newcolumntype{C}{>{\centering\arraybackslash}X}
\begin{tabularx}
{\textwidth}{CCCC}
\toprule
 & $E_T$ & $E_{q_x}$ & $E_{q_y}$\\
\midrule
$\tau_\mathrm{ref} = 0.1$ & 1.605\% & 10.283\% & 7.337\%\\
$\tau_\mathrm{ref} = 1.0$ & 1.170\% & 5.448\% & 3.187\%\\
\bottomrule
\end{tabularx}
\end{table}

Figures~\ref{fig:9a} and \ref{fig:9e} show that the inferred $\tau(x,y)$ agrees well with the reference in the interior but deviates near the domain boundaries.
The reconstructed temperature and heat flux fields are compared in Figs.~\ref{fig:9b}--\ref{fig:9d} and \ref{fig:9f}--\ref{fig:9h}, with quantitative errors summarized in Tables~\ref{tab:tab4} and \ref{tab:tab5}.
Although the relative errors in $\tau$ are substantial (21\%--33\%) due to localized boundary deviations, the reconstructed temperature fields maintain high accuracy, with relative $L_2$ errors below $2\%$ in both cases.
The heat fluxes $q_x$ and $q_y$ exhibit larger relative errors, primarily because their small magnitudes amplify relative discrepancies, though the overall trends and distributions are captured reasonably well.
Numerical simulations using the inferred $\tau(x,y)$ reproduce the macroscopic temperature and heat flux fields with moderate accuracy (Table~\ref{tab:tab5}), confirming that the inferred relaxation time captures the dominant thermal response despite local inaccuracies.

The larger boundary errors in $\tau$ can be attributed to two factors.
First, the sensor points are distributed in the interior, leaving the boundaries without direct observational constraints.
Second, as discussed in~\ref{sec:appendix_b}, the phonon energy density $e^{\prime\prime}$ may exhibit angular discontinuities at boundaries under thermalization conditions, which complicates the inverse inference of local $\tau$ from the governing-equation residual.

\section{Summary}\label{sec:conclusion}
We have extended the Monte Carlo physics-informed neural networks (MC-PINNs) developed for solving the phonon Boltzmann transport equation in \cite{lin2025monte} to solve inverse multiscale heat conduction problems governed by the phonon Boltzmann transport equation.
By embedding the phonon BTE directly into the neural networks, the MC-PINN reconstructs thermal fields and identifies unknown thermophysical parameters from limited measurements while maintaining consistency with the governing transport physics.
The mesh-free Monte Carlo sampling strategy, inherited from the forward MC-PINNs formulation~\cite{lin2025monte}, enables a unified treatment across diffusive, transitional, and ballistic transport regimes without requiring a priori knowledge of the relaxation time.

Two representative classes of inverse problems were investigated.
For thermal field reconstruction under unknown boundary conditions, MC-PINNs consistently outperformed purely data-driven DNNs across quasi-2D and 3D benchmarks spanning a wide range of Knudsen numbers.
The advantage is most pronounced in regions where DNNs struggle to extrapolate from sparse interior measurements, particularly near heat sources with localized thermal gradients and near unobserved boundaries. These discrepancies are especially evident for the transitional-regime FinFET case ($\tau=100.0$).
The BTE residual effectively propagates information from the sensor locations along the characteristic directions of the kinetic equation, compensating for the absence of boundary data and enabling accurate reconstruction even in device-scale 3D geometries with geometric complexity.
For simultaneous inference of the thermal field and unknown relaxation time, the framework accurately identified spatially uniform $\tau$ with relative errors below $1\%$ in both quasi-1D and quasi-2D settings, even with as few as two sensor points.
When $\tau$ was treated as a spatially varying function and modeled by a sub-network, the inferred distributions captured the dominant thermal response: numerical simulations using the recovered $\tau(\boldsymbol{x})$ reproduced the macroscopic temperature and heat flux fields with good accuracy, despite localized pointwise deviations in $\tau$ near boundaries and in regimes of weak physical sensitivity.

The results for spatially varying $\tau$ highlight a broader open problem: the robust pointwise identification of spatially heterogeneous transport coefficients from sparse macroscopic observables.
The difficulty is fundamental rather than architectural: the mapping from $\tau(\boldsymbol{x})$ to $(T, \boldsymbol{q})$ is non-injective and ill-conditioned, especially near the ballistic and diffusive limits where the macroscopic fields become insensitive to variations in $\tau$.
Addressing this challenge motivates the integration of complementary data sources, such as surface measurements of the distribution function, multi-frequency observations, or material priors, into the MC-PINNs.
More broadly, the present study shows that MC-PINNs can serve as a physically consistent framework for inverse thermal analysis at micro- and nanoscales, with potential applications in thermal imaging, hotspot detection, material characterization, and temperature monitoring in microelectronic devices.

\section*{Acknowledgments}
Q. L., X. M., and Z. G. acknowledge the support of the National Natural Science Foundation of China (No. 12201229) and the Interdisciplinary Research Program of HUST (No. 2024JCYJ003 and 2023JCYJ002).
C. Z. acknowledges the support of the National Natural Science Foundation of China (52506078) and Zhejiang Provincial Natural Science Foundation of China under Grant No.~LMS26E060012.
X. M. also acknowledges the support of the Xiaomi Young Talents Program. The authors acknowledge Beijing PARATERA Tech. CO., Ltd. for the HPC resources.

\appendix
\setcounter{figure}{0}
\setcounter{table}{0}
\section{Computational Details}
\label{sec:appendix_c}

In all test cases presented in Sec.~\ref{sec:results}, the Adam optimizer with a learning rate of $10^{-3}$ is employed for training MC-PINNs.
The detailed configurations are summarized in Table~\ref{app._tab1}.
For quasi-2D cases, 128 points are used for Monte Carlo integration of the equilibrium energy density, which is found to be sufficient for accurate quadrature.

\begin{table}[H]
\caption{Network architectures and sensor points used by MC-PINNs in each case.}
\label{app._tab1}
\newcolumntype{C}{>{\centering\arraybackslash}X}
\begin{tabularx}
{\textwidth}{m{3cm}m{2.5cm}m{3cm}m{3cm}m{3cm}}
\toprule
 Section & & Width $\times$ Depth & Activation function & Sensor points ($N_{\mathrm{sen}}$) \\
\midrule
\ref{sec:results_known_tau} (quasi-2D) & Uniform $\tau$  & $30\times 4 $ & tanh & 20 \\
  & Varying $\tau$ & $30\times 4 $ & tanh & 20 \\
\ref{sec:results_known_tau} (3D) & FinFET & $50\times 4 $ & tanh & 200 \\
\ref{sec:results_unknown_tau} (uniform) & Quasi-1D & $20\times 3 $ & tanh & 2 \\
  & Quasi-2D & $30\times 4 $ & tanh & 40 \\
\ref{sec:results_unknown_tau} (varying) & Quasi-1D & $20\times 3,\,20\times 2 $ & tanh, exp & 40 \\
  & Quasi-2D & $50\times 4 ,\,30\times 3 $ & tanh, exp & 40 \\
\bottomrule
\end{tabularx}
\end{table}

Details of the numerical grids used to generate the reference solutions are provided in Table~\ref{app._tab2}.
Uniform spatial grids are adopted in all cases, with Gauss--Legendre quadrature applied in the solid angular domain.

\begin{table}[H]
\caption{Grid sizes used by the numerical methods to generate reference solutions.}
\label{app._tab2}
\newcolumntype{C}{>{\centering\arraybackslash}X}
\begin{tabularx}
{\textwidth}{m{3cm}m{3cm}m{4cm}m{4.5cm}}
\toprule
 Section & & Space ($\boldsymbol{x}$) & Solid angular domain ($\boldsymbol{s}$) \\
\midrule
\ref{sec:results_known_tau} (quasi-2D) & Uniform $\tau$~\cite{zhang2023acceleration} & $100\times 100$,\,$40\times 40$ & $32\times 16$,\,$100\times 100$ \\
  & Varying $\tau$~\cite{zhang2023acceleration} & $80\times 80$ & $100\times 100$ \\
\ref{sec:results_known_tau} (3D) & FinFET~\cite{zhang2025effects} & $dx=dy=1,\, dz=2$ & $48\times 48$ \\
\ref{sec:results_unknown_tau} (uniform) & Quasi-1D~\cite{zhang2019implicit} & 40 & 100 \\
  & Quasi-2D~\cite{zhang2023acceleration} & $40\times 40$ & $100\times 100$ \\
\ref{sec:results_unknown_tau} (varying) & Quasi-1D~\cite{zhang2019implicit} & 40 & 100 \\
  & Quasi-2D~\cite{zhang2023acceleration} & $40\times 40$ & $100\times 100$ \\
\bottomrule
\end{tabularx}
\end{table}

\setcounter{figure}{0}
\section{Sensitivity Analysis of the Relaxation Time}\label{sec:appendix_a}

To elucidate the challenges encountered in identifying spatially varying relaxation times (Sec.~\ref{sec:results_unknown_tau}), we analyze the sensitivity of the macroscopic temperature field to perturbations in $\tau$ using the quasi-1D benchmark configuration.
Two sets of comparisons are designed to examine the system response near the ballistic and diffusive limits.
In the first set, the baseline relaxation times are $\tau_{\max}=10.0$ (left) and $\tau_{\min}=0.1$ (right). With $\tau_{\min}$ fixed, $\tau_{\max}$ is increased by $50\%$ to assess the influence of large $\tau$ on the temperature distribution.
In the second set, the baseline values are $\tau_{\min}=0.01$ and $\tau_{\max}=1.0$. Here, $\tau_{\max}$ is fixed while $\tau_{\min}$ is reduced by $50\%$ to assess the sensitivity to small relaxation times.

\begin{figure}[H]
\centering
\subfigure[]{\includegraphics[width=0.48\textwidth]{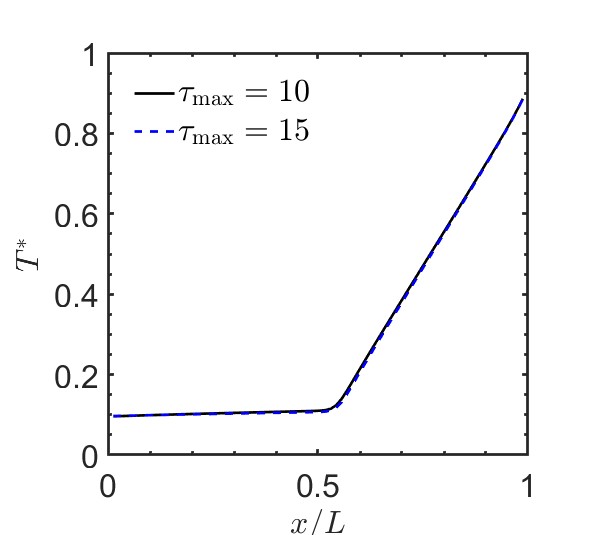}\label{fig:app_T1_1}}
\subfigure[]{\includegraphics[width=0.48\textwidth]{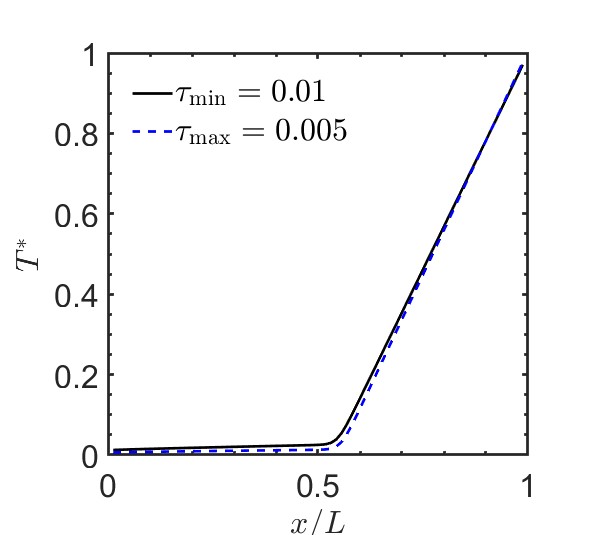}\label{fig:app_T1_2}}
\caption{
Sensitivity of the temperature field to variations in the relaxation time $\tau$.
(a) Effect of increasing $\tau_{\max}$ by $50\%$ with fixed $\tau_{\min}$.
(b) Effect of decreasing $\tau_{\min}$ by $50\%$ with fixed $\tau_{\max}$.}
\label{fig:app_T1}
\end{figure}

Figure~\ref{fig:app_T1} presents the resulting temperature distributions.
As shown in Fig.~\ref{fig:app_T1}(a), when $\tau$ is already large ($\tau>10$, approaching the ballistic regime), a further increase produces a negligible change in the temperature field.
Similarly, Fig.~\ref{fig:app_T1}(b) shows that when $\tau$ is sufficiently small ($\tau<0.01$, approaching the diffusive regime), further reduction has minimal impact.
In both limits, the macroscopic field is primarily governed by the boundary constraints and is therefore insensitive to $\tau$ over a relatively wide range.
This weak sensitivity reduces identifiability from macroscopic data and can lead to non-unique or inaccurate parameter estimates.

To further assess sensitivity in the transitional regime ($\tau \in [0.1,\,1.0]$), we consider a baseline with $\tau_{\min}=0.1$ and $\tau_{\max}=1.0$, and introduce $\pm50\%$ perturbations at both ends.

\begin{figure}[H]
\centering
\subfigure[]{\includegraphics[width=0.48\textwidth]{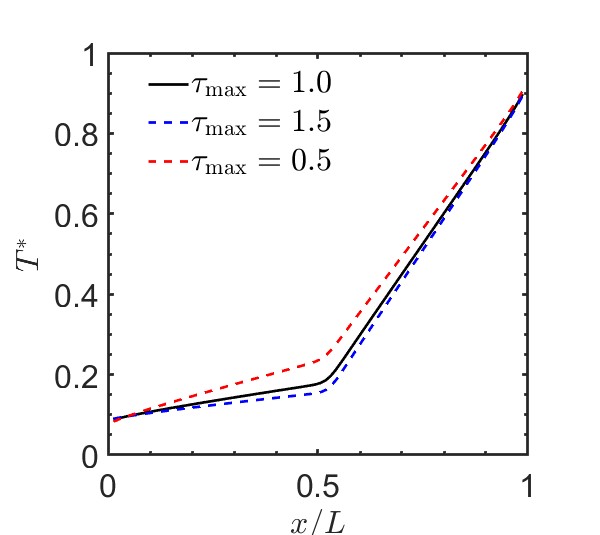}\label{fig:app_T2_1}}
\subfigure[]{\includegraphics[width=0.48\textwidth]{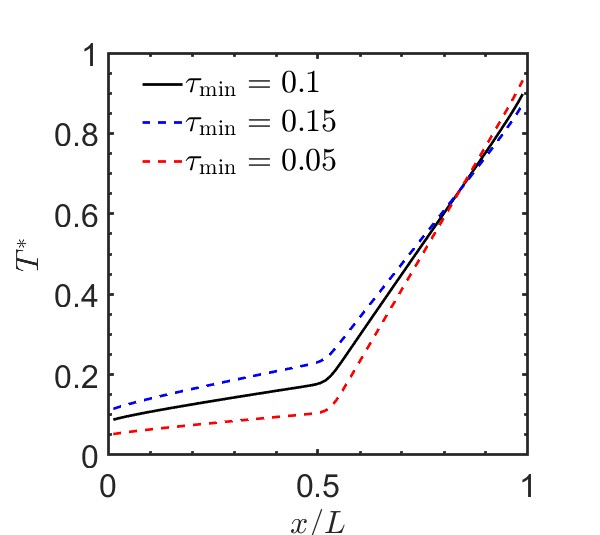}\label{fig:app_T2_2}}
\caption{
Sensitivity of the temperature field to variations in $\tau$ within $\tau \in [0.1,\,1.0]$.
(a) Effect of $\pm50\%$ perturbation in $\tau_{\max}$ with fixed $\tau_{\min}$.
(b) Effect of $\pm50\%$ perturbation in $\tau_{\min}$ with fixed $\tau_{\max}$.}
\label{fig:app_T2}
\end{figure}

Figures~\ref{fig:app_T2}(a) and \ref{fig:app_T2}(b) show that perturbations in $\tau$ within the transitional regime produce more noticeable changes in the temperature field compared with the limiting regimes.
Moreover, changes in the smaller relaxation time exert a stronger influence than those in the larger one, consistent with the $1/\tau$ scaling of the collision frequency in the scattering term.
For large $\tau$, the collision frequency is already low, and further changes produce relatively weak modifications of the scattering process.
Consequently, the macroscopic field is inherently less sensitive to perturbations in regions with large $\tau$, making accurate estimation in such regions more difficult. This is consistent with the inversion results in Fig.~\ref{fig:8a}, where larger deviations from the reference occur in the high-$\tau$ region.
The weak sensitivity implies that multiple relaxation-time distributions may be compatible with the same observed macroscopic temperature field, which is an inherent limitation of inferring spatially varying transport coefficients from sparse, macroscopic data alone.

\setcounter{figure}{0}
\section{Discontinuity of the Phonon Energy Density at Boundaries}\label{sec:appendix_b}

Under thermalization boundary conditions, the phonon energy density $e^{\prime\prime}$ may exhibit discontinuities at solid boundaries.
The incoming phonon distribution at the wall is prescribed as the equilibrium distribution at the wall temperature, while the outgoing distribution is determined by the solution of the BTE inside the domain, naturally introducing angular discontinuities.

To illustrate this effect, we compare $e^{\prime\prime}$ near the top boundary for the quasi-2D heat conduction problem discussed in Sec.~\ref{sec:results_unknown_tau}.

\begin{figure}[H]
\centering
\subfigure[]{\includegraphics[width=0.32\textwidth]{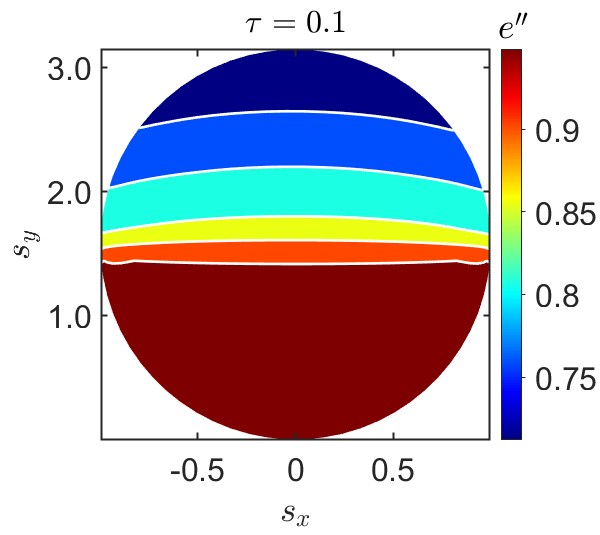}\label{fig:app_f1}}
\subfigure[]{\includegraphics[width=0.32\textwidth]{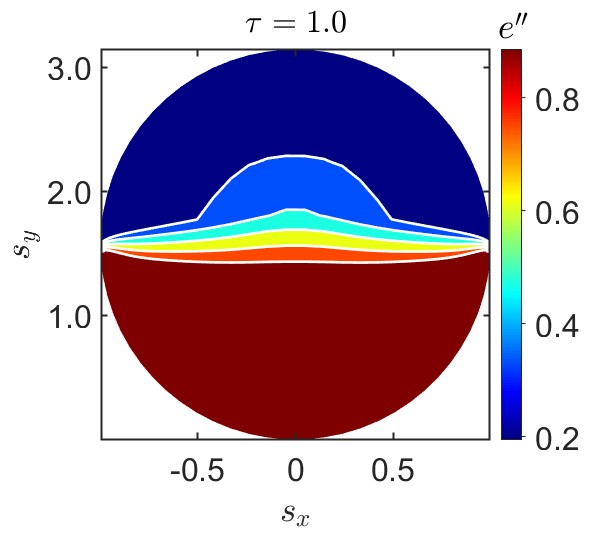}\label{fig:app_f2}}
\subfigure[]{\includegraphics[width=0.32\textwidth]{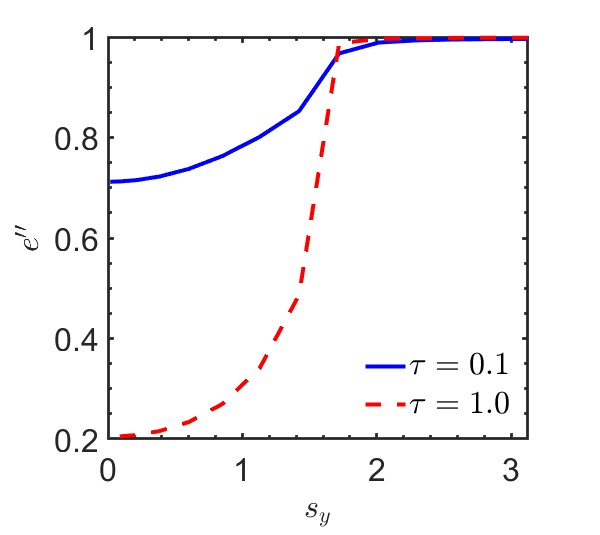}\label{fig:app_f_line}}
\caption{
Phonon energy density $e^{\prime\prime}$ in angular space at the top boundary center $(x=0.5,\,y=1.0)$: (a) $\tau = 0.1$, (b) $\tau = 1.0$, and (c) angular profiles along $s_y$ at $s_x=0.0$ for both cases.}
\label{fig:app_f}
\end{figure}

Figure~\ref{fig:app_f} displays the angular distribution of $e^{\prime\prime}$ at the center of the top wall $(x=0.5,\,y=1.0)$.
For $\tau=0.1$ (near-diffusive, Fig.~\ref{fig:app_f}(a)), the distribution remains relatively smooth, as scattering is still dominant.
For $\tau=1.0$ (transitional, Fig.~\ref{fig:app_f}(b)), ballistic effects become pronounced, and a clear discontinuity appears between the imposed incoming distribution and the computed outgoing distribution.
This contrast is further illustrated in Fig.~\ref{fig:app_f}(c), which shows the density profiles along $s_y$ at $s_x=0.0$ for both cases.

Since MC-PINNs approximate the solution using smooth neural networks, such sharp boundary discontinuities can induce larger local approximation errors, which propagate into the governing-equation residual and reduce the accuracy of $\tau$ recovered from the residual at the boundary.
This provides a plausible explanation for the larger boundary errors observed in the inferred $\tau(x,y)$ in Figs.~\ref{fig:9a} and \ref{fig:9e} of the main text.

\bibliographystyle{elsarticle-num}
\bibliography{ref}

\end{document}